\documentclass[prd, amsfonts, twocolumn, nofootinbib, showpacs]{revtex4}
\usepackage{graphicx, epsfig}
\usepackage{color}
\usepackage{amsmath}
\usepackage{hyperref}
\usepackage{subfig}
\usepackage{tabularx}
\newcommand{\be}{\begin{equation}}
\newcommand{\ee}{\end{equation}}
\newcommand{\bea}{\begin{eqnarray}}
\newcommand{\eea}{\end{eqnarray}}

\newcommand{\gapp}{\mathrel{\raise.3ex\hbox{$>$}\mkern-14mu \lower0.6ex\hbox{$\sim$}}}
\newcommand{\lapp}{\mathrel{\raise.3ex\hbox{$<$}\mkern-14mu \lower0.6ex\hbox{$\sim$}}}
\def\bbox{{\,\lower0.9pt\vbox{\hrule \hbox{\vrule height 0.2 cm
\hskip 0.2 cm \vrule  height 0.2 cm}\hrule}\,}}

\usepackage{color}
\usepackage{ulem}
\usepackage[usenames,dvipsnames]{xcolor}

\begin{document}
\title{Greybody factors for a black hole in massive gravity}
\author{Ruifeng Dong, Dejan Stojkovic}
\affiliation{ HEPCOS, Department of Physics, SUNY at Buffalo, Buffalo, NY 14260-1500}

\begin{abstract}
\widetext
An exact solution was recently found in the massive gravity theory having the form of Schwarzschild-dS black holes with some additional background fields. Hawking radiation will occur at the event and cosmological horizons having the blackbody spectrum, which will be modified by the geometry outside the black hole.
In this paper, we study the greybody factors of a test scalar, considering its minimal coupling with the background geometry. The case of small black holes with the horizon radius much smaller than the cosmological dS radius is studied numerically.   The case of near-extremal black holes with the horizon radius comparable to the cosmological dS radius is studied analytically.
In addition, we considered the coupling of the test field with the background {St\"uckelberg} fields, which in turn leads to reductions in particle emission and some non-trivial features (resonances) in the greybody factors.
\end{abstract}


\maketitle

\section{Introduction}

Hawking emission from a black hole is characterized by two factors: the black body factor which gives the probability that a particle is created in the vicinity of a horizon, and the greybody factor which gives the probability that this particle penetrates the potential barrier and escapes to infinity. The pedagogical difference between these two factors can be found in \cite{Dai:2010xp} (see also \cite{Brito:2015oca}). The number of distinct models with the exact black hole solutions is constantly increasing, which in turn implies the need for concrete calculations of the corresponding greybody factors (for some recent developments see for example
\cite{Dai:2007ki,Dai:2009by,Dai:2008qn,Dai:2006hf,Frolov:2002xf,Frolov:2002gf,Frolov:2002as,Kanti:2014vsa,Kanti:2010mk,Kanti:2009sn,Casals:2008pq,Creek:2006ia,Kanti:2002nr}).
The main goal in this paper is to calculate the greybody factors for black holes in massive gravity, which is the case that has not been studied in the literature so far.

Massive gravity is a collective name for theories where gravitons appear to be massive. The original idea is old, appearing for the first time in \cite{FP}, but gained a lot of attention recently.   Namely, massive gravitons emerge naturally in some higher-dimensional models like the Dvali, Gabadadze, Porrati model \cite{Dvali:2000hr}, cascading gravity \cite{rvhk,rkt,rhkt,Dai:2014roa,Hao:2014tsa,Kaloper:2007ap,Kaloper:2007qh} and ghost-free massive gravity \cite{deRham:2010kj}. For nice reviews please see \cite{Hinterbichler:2011tt,deRham:2014zqa}.

In massive gravity models, the asymptotically flat solution of vacuum Einstein equations converts the horizon to a spacetime singularity. One way out of this problem is to study the asymptotically de Sitter solutions \cite{Berezhiani:2011mt}. Then the simplest case is a Schwarzschild-dS black hole,
with background St\"uckelberg scalar fields. As a first study of Hawking emission in this class of black holes, we consider the emitted scalar field minimally coupled to the gravitational background with and without coupling to these external fields. Though emission from a Schwarzschild-dS black hole has been studied before, coupling to the background fields will introduce some non-trivial features in the greybody factors.

The Lagrangian we will consider is
\be
\mathcal{L}= \sqrt{-g}\left( \frac{M^2_{pl}}2(R+{m_g}^2 \mathcal{U}(g,\phi^a)) -\frac12 g^{\mu\nu}\partial_{\mu}\varphi\partial_{\nu}\varphi \right).
\ee
Here $m_g$ is the graviton mass and $\mathcal{U}$ is the graviton potential which depends on the metric as well as the St\"uckelberg fields $\phi^a$. Explicitly,
\be
\mathcal{U}(g,\phi^a)=\mathcal{U}_2 + \frac{\alpha-1}{3}\mathcal{U}_3 -\left(\frac{\eta}{2}+\frac{\alpha-1}{12}\right)\mathcal{U}_4,
\ee
where $\alpha_3$, $\alpha_4$ are free parameters, and
\bea
\mathcal{U}_2&= &[\mathcal{K}]^2-[\mathcal{K}^2], \nonumber \\
\mathcal{U}_3&= &[\mathcal{K}]^3-3[\mathcal{K}][\mathcal{K}^2]+2[\mathcal{K}^3], \nonumber \\
\mathcal{U}_4&= &[\mathcal{K}]^4-6[\mathcal{K}^2][\mathcal{K}]^2+8[\mathcal{K}^3][\mathcal{K}]+3[\mathcal{K}^2]^2-6[\mathcal{K}^4],  \nonumber\\
\eea
with the matrix $\mathcal{K}^{\mu}_{\nu}(g,\phi^a)=\delta^{\mu}_{\nu}-\sqrt{g^{\mu\alpha}\partial_{\alpha}\phi^a\partial_{\nu}\phi^b\eta_{ab}}$, and $[\mathcal{K}]$ is the trace of it.

 As found in \cite{Berezhiani:2011mt}, neglecting the backreaction of the test fields $\varphi$ on spacetime, there exist Schwarzschild-dS type of black hole solutions for a special choice of
parameters $\alpha$ and $\eta$ as $\eta=-\frac{\alpha^2}{6}$.
\be
ds^2=-(1-\frac{r_g}r-\frac2{3\alpha}{m_g}^2 r^2)dt^2+\frac{dr^2}{1-\frac{r_g}r-\frac2{3\alpha}{m_g}^2 r^2}+r^2d\Omega^2,
\label{ds2}
\ee
where the free parameter $r_g$ is the gravitational radius of the black hole. In this solution, the cosmological horizon size is determined by the the graviton mass. In what follows,
we consider the equation of motion of $\varphi$ in this given spacetime.

\section{General analysis}

Now we consider the spherically symmetric spacetime,
\be
ds^2 = -F(r) dt^2 +F^{-1} (r) dr^2 +r^2 d\Omega^2.
\ee
For the scalar Hawking emission, the equation governing its kinematics is the Klein-Gordon equation. The scalar field must retain the form where the variables can be separated as
\be
\varphi(t,r,\Omega)=\frac1{r}e^{i\omega t}\Psi_{\omega,l}(r) Y_{lm}(\Omega),
\label{seperate}
\ee
where $\omega$ is the (real) frequency under consideration, and $Y_{lm}$ are the spherical harmonics. Now the Klein-Gordon equation reduces to a single radial equation,
\be \label{kge}
\left[\frac{d^2}{d{r^*}^2}+\omega^2-V(r) \right]\Psi_{\omega,l}=0.
\ee
Here $r^*$ is the tortoise coordinate defined as
\be
r^*=\int \frac{dr}{F(r)},
\ee
and $V(r)$ is the effective potential determined by the metric,
\be
V(r)=\frac{F(r)\partial_r F(r)}{r} + l(l+1)\frac{F(r)}{r^2}.
\label{V(r)}
\ee

We now rewrite $F(r)$ concretely for Schwarzschild-dS black holes (\ref{ds2}) in the popular form,
\be
F(r)=1-r_g/r-\kappa^2 r^2,
\ee
with $\kappa\equiv \sqrt{\frac2{3\alpha}}m_g$. The event horizon $R_H$ and the cosmological horizon $R_C$ will be two of the three roots of $F(r)$, and the third one denoted as $R_X$, i.e.
\be
F(r)=-\frac{\kappa^2(r-R_H)(r-R_C)(r-R_X)}{r}.
\ee
By extracting out the constant term in the numerator, we get
\be
r_g=-R_H R_C R_X > 0,
\ee
thus $R_X=-\frac{r_g}{R_H R_C}<0$, which means that there are no singularities,  coordinate or physical, between the event and cosmological horizons, which makes it straightforward to
 numerically solve Eq. (\ref{kge}) between $R_H$ and $R_C$.

In terms of the parameters in massive gravity,
\bea
R_H &\approx&r_g, \\
R_C &\approx&\frac{\sqrt{3\alpha/2}}{m_g},
\eea
for small black holes $R_H\ll R_C$, and
\be
R_H\approx R_C\approx \frac{\sqrt{\alpha/2}}{m_g}\approx \frac{3 r_g}2,
\ee
for near-extremal black holes $R_H\approx R_C$.

We can see that $F(r)$ vanishes at both event and cosmological horizons, and that $r^*\rightarrow -\infty$ as $r\rightarrow R_H$, $r^*\rightarrow +\infty$ as $r\rightarrow R_C$. Therefore, the
solution to Eq. (\ref{kge}) is readily written once the boundary condition is specified.

First, if particles are emitted from the event horizon, travel to the cosmological horizon and are partly reflected back,
then the solution ("up" modes) assumes the following form,
\bea
\Psi &\sim& e^{-i\omega r^*} + R e^{i\omega r^*}, r^*\rightarrow -\infty \nonumber \\
&\sim& T e^{-i\omega r^*}, r^*\rightarrow +\infty.
\label{eqn1}
\eea
Second, if particles are injected from the cosmological horizon, partly reach the event horizon and partly get reflected back, then the solution ("in" modes) should have the form
\bea
\Psi &\sim& T' e^{i\omega r^*}, r^*\rightarrow -\infty \nonumber \\
&\sim& e^{i\omega r^*} + R' e^{-i\omega r^*}, r^*\rightarrow +\infty.
\label{eqn2}
\eea
For real frequencies, the radial equation is invariant under complex conjugation, so there is another solution,
\bea
\Psi &\sim& e^{i\omega r^*} + R^* e^{-i\omega r^*}, r^*\rightarrow -\infty \nonumber \\
&\sim& T^* e^{i\omega r^*}, r^*\rightarrow +\infty.
\label{eqn3}
\eea
A linear combination of solutions (\ref{eqn1}) and (\ref{eqn3}) is also a solution of the linear equation (\ref{kge}), so we can define
\bea
\Psi &\sim& (1-RR^*)e^{i\omega r^*}, r^*\rightarrow -\infty \nonumber \\
&\sim& T^* e^{i\omega r^*} - R^* T e^{-i\omega r^*}, r^*\rightarrow +\infty.
\label{eqn4}
\eea
Equations~(\ref{eqn2}) and (\ref{eqn4}) are two linearly dependent solutions, thus
\be
\frac{T'}{1-RR^*}=\frac1{T^*}=-\frac{R'}{R^* T},
\ee
Combining this with the definition of the greybody factors $\gamma(\omega)=TT^*=1-RR^*$, we get
\be
TT^*=T'T'^*.
\ee
This means that the black hole particle transmission coefficient (greybody factor) is equal to the transmission coefficient for emissions from the cosmological horizon.

\begin{figure}[h]
 \begin{center}
\includegraphics[width=2.0in]{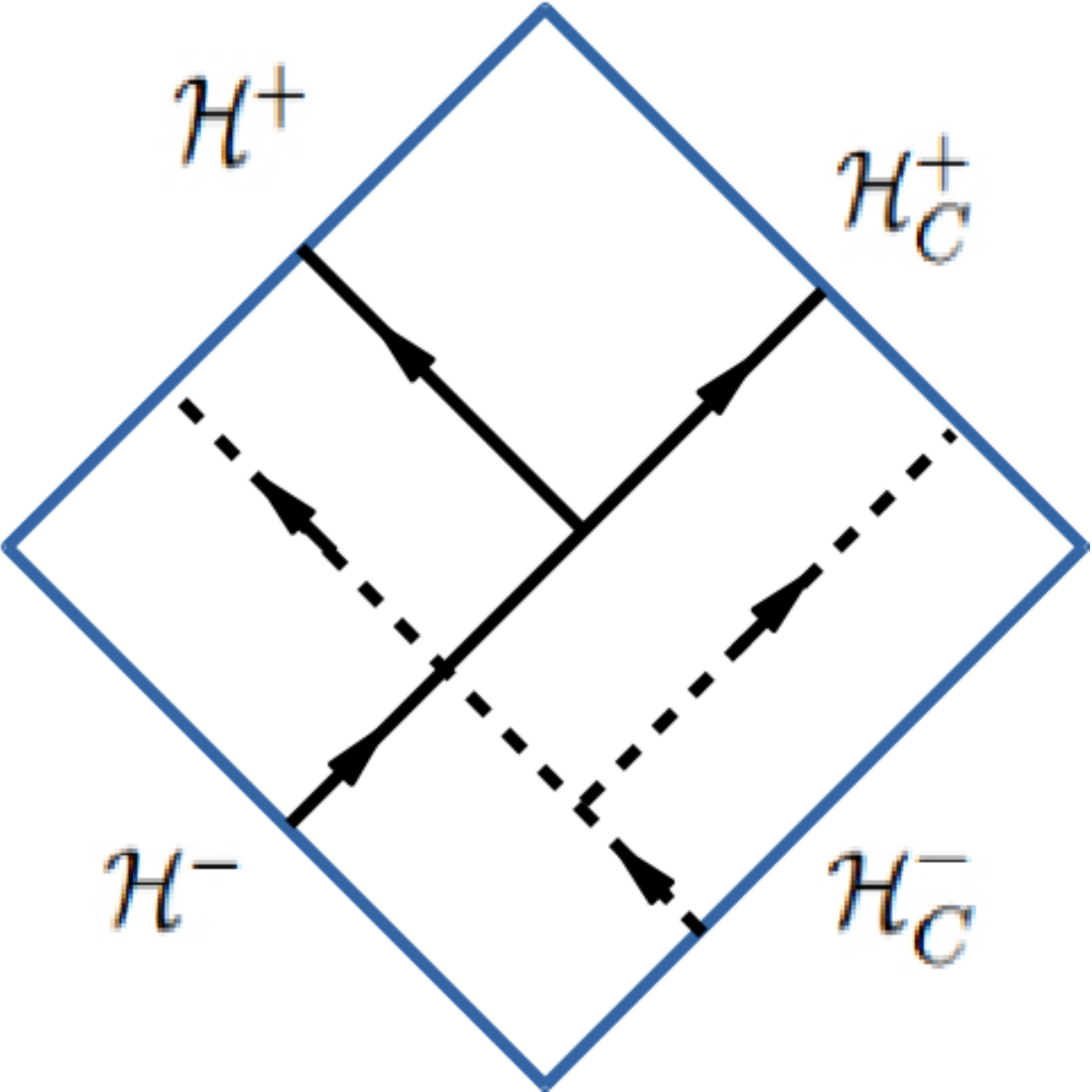}
\caption{Partial Penrose diagram for the Schwarzschild-dS spacetime, along with the emission and transmission of the emitted particles from both horizons.\cite{Harmark:2007jy}}
\label{fig-penrose}
\end{center}
\end{figure}

In Penrose diagram, Fig.~(\ref{fig-penrose}) \cite{Harmark:2007jy}, particles emitted from the past event horizon $\mathcal{H}^-$ partly travel to the future cosmological
 horizon $\mathcal{H}^+_C$ and partly get reflected back to the future event horizon $\mathcal{H}^+$. Similarly, particles emitted from the past cosmological horizon $\mathcal{H}^-_C$
 are partly received at the future event horizon $\mathcal{H}^+$, and partly reflected to the future cosmological horizon $\mathcal{H}^+_C$. And as shown above, the greybody factors
 for these two processes are equal.

\section{Numerical results}

\begin{figure}[h]
   \centering
\includegraphics[height=0.35\textwidth,angle=0]{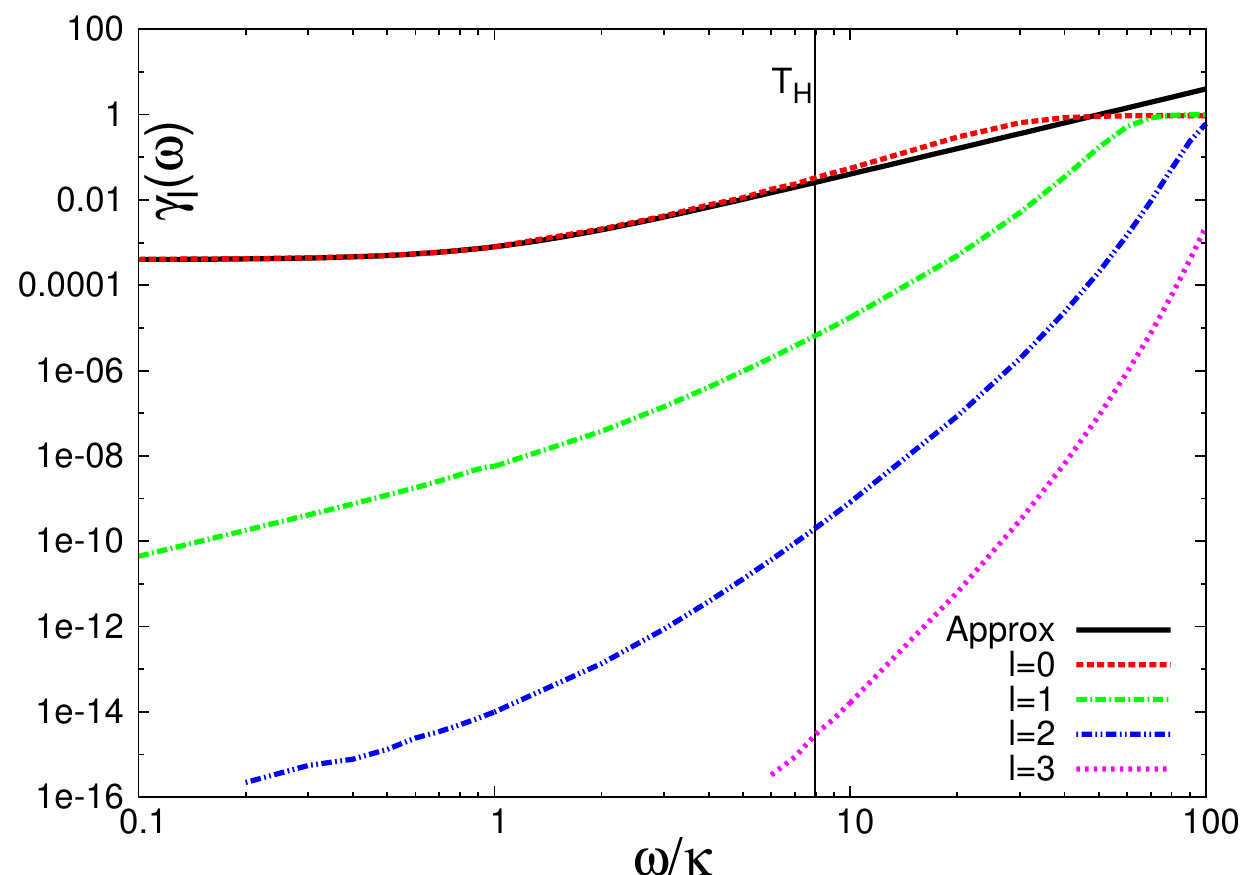}
\caption{Greybody factors for Schwarzschild-dS black holes for $l$ from $0$ to $3$. The thin vertical black line shows the position of the Hawking temperature, and the thick solid black line is the low-frequency appximation in Eq. (\ref{approx_low_w}).}
\label{fig-g_l}
\end{figure}

\begin{figure}[h]
  \centering
\includegraphics[height=0.35\textwidth,angle=0]{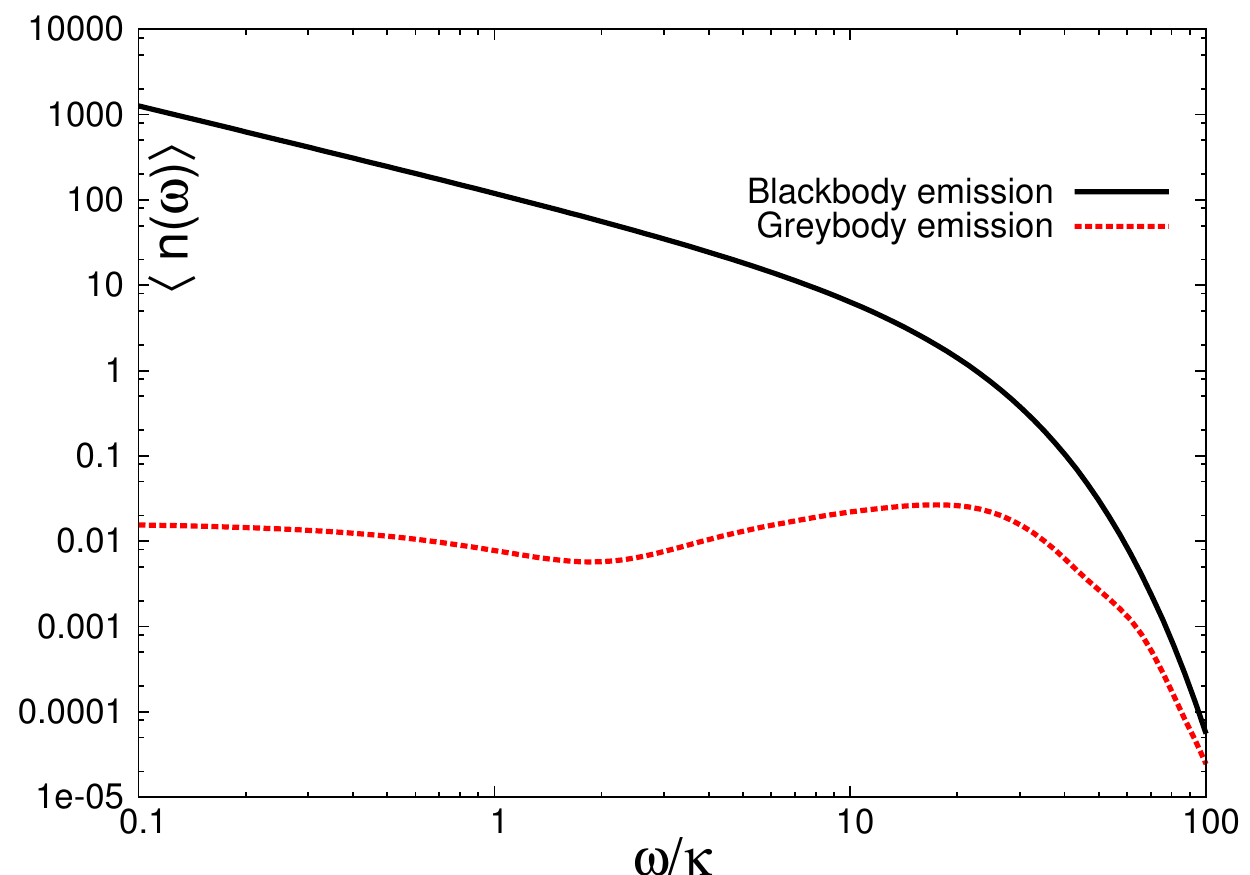}
\caption{The expected emission rate vs. frequency, including modes with $l=0,1,2,3$, for small Schwarzschild-dS black holes. }
\label{fig-n_w}
\end{figure}

\begin{figure}[h]
  \centering
\includegraphics[height=0.35\textwidth,angle=0]{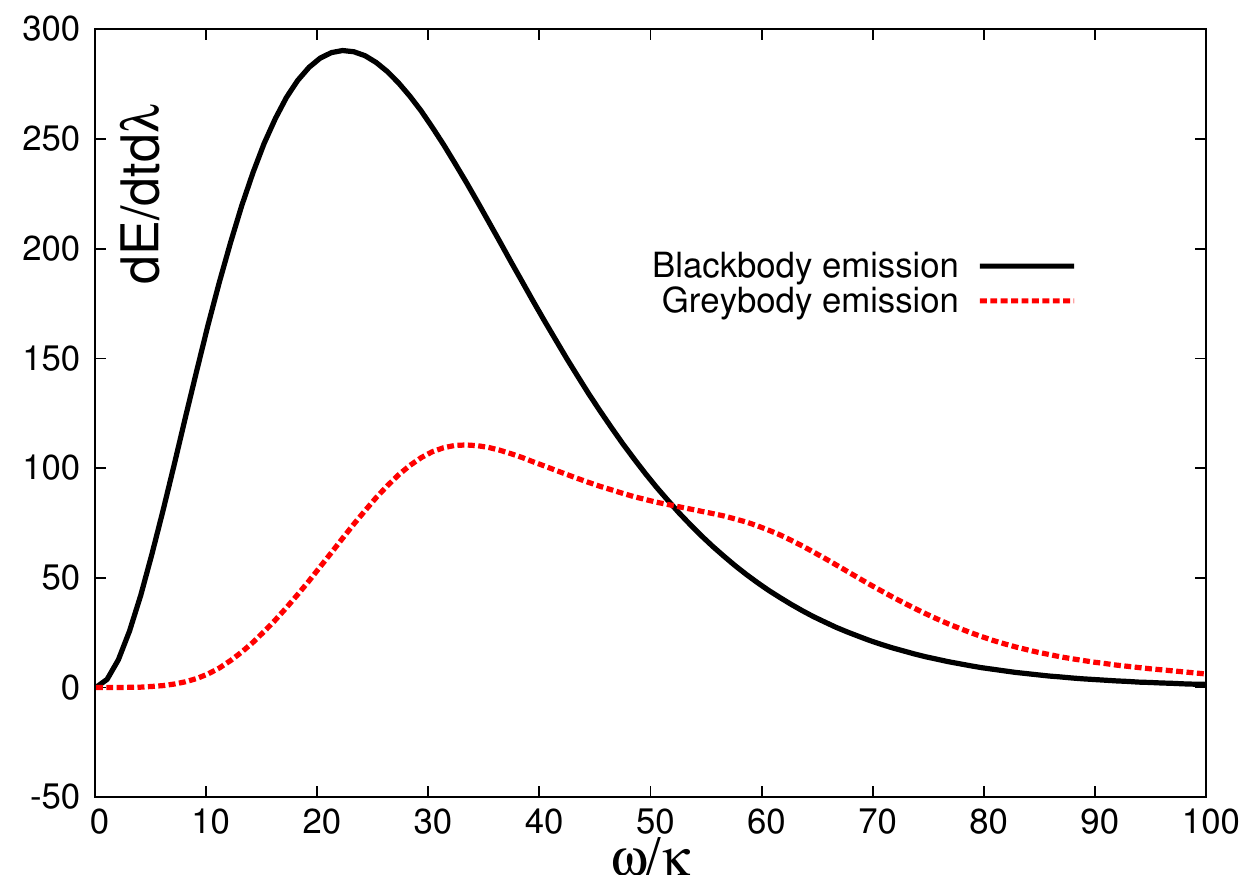}
\caption{Emission power spectrum of small Schwarzschild-dS black holes, compared with the blackbody power spectrum. Here $\lambda\equiv 2\pi / \omega$, where we have taken the speed of light to be 1. The greybody spectrum is amplified by a factor of $10$ to be more visible. $dE/dt d\lambda$ has the unit $\kappa^3$.
Note that the greybody spectrum is always below the blackbody one before the artificial amplification.}
\label{fig-power}
\end{figure}

In this section we restrict our attention to small black holes, that is, $R_H\ll R_C$ or $M\ll 1/\kappa$. We chose $\kappa R_H=0.01$ without loss of generality. For this horizon radius, the surface temperature $T_H=\frac{k_H}{2\pi}=\frac{F'(R_H)}{4\pi}\approx 8\kappa$.
To obtain the black hole emission rate and spectrum, we take the cutoff $l$ to be 3. We adopt the fourth-order Runge-Kutta method in the numerical integration. To get convergent results independent of the computation,
we chose the step size of integration as $10^{-3}$ for $l=0, 1, 2$, and $10^{-4}$ for $l=3$ to treat the higher frequency modes correctly. The results are shown in the figures (\ref{fig-g_l})(\ref{fig-n_w})(\ref{fig-power}).

As seen from equation (\ref{V(r)}), the potential barrier gets higher as $l$ gets larger, making it more difficult for low-energy particles to get across, consistent with Fig. (\ref{fig-g_l}). Besides, the contribution to the total emission rate and power from modes with frequencies well beyong the
Hawking temperature is exponentially depressed. So we restrict ourselves to $\omega/\kappa \le 100$, at which energy the mode $l=3$ contributes less than $1/1000$, as seen from Fig. (\ref{fig-g_l}). This gives roughly a precision of one part in a
thousand in the spectral emission rate and power spectrum we obtained in figures (\ref{fig-n_w})(\ref{fig-power}).

Our result for $l=0$ at low frequencies is consistent with the analytical approximation of Harmark et al. \cite{Harmark:2007jy},
\be
\gamma_0(\omega)\approx 4(\kappa R_H)^2(1+\omega^2/\kappa^2),
\label{approx_low_w}
\ee
as was derived by matching the solution forms in different regions of $r$, and is also consistent with the result of greybody factors in the $\omega\rightarrow 0$ limit for $(4+n)$-dimensional Schwarzschild-dS black holes by Kanti et al. \cite{Kanti:2005ja},
\be
\gamma_{0,n}(\omega\rightarrow 0) = 4\frac{(R_C R_H)^{n+2}}{(R_C^{n+2}+R_H^{n+2})^2},
\ee
in the case of $n=0$ and $R_H\ll R_C$. Qualitatively, our numerical results also agree with those by Crispino et al. \cite{Crispino:2013pya}, though for a different small event horizon size.

We also plot the expected emission rate $\langle n(\omega)\rangle$ and the power spectrum $\frac{dE}{dtd\lambda}$, which have the well-known expressions as
\bea
\langle n(\omega)\rangle &=& \sum_l\frac{(2l+1)\gamma_l(\omega)}{\exp(2\pi\omega/k_H)-1} \\
\frac{dE}{dtd\lambda}&=&\frac{\omega^2}{2\pi}\frac{dE}{dtd\omega}=\frac{\omega^3}{4\pi^2}\sum_l \frac{(2l+1)\gamma_l(\omega)}{\exp(2\pi\omega/k_H)-1}. \nonumber\\
\eea
Compared with the blackbody radiation, the emission rate is largely reduced at low frequencies due to the small greybody factors there. This reduction shrinks at higher frequencies. Also, the peak frequency on the power spectrum  moves to higher values, from that of the Planck's radiation specturm.

\section{Near-extremal black holes}

 In this section we consider the case of near-extremal Schwarzschild-dS black holes, where the black hole event horizon and cosmological horizon are very close. Now we can get approximately a closed form of $V(x)$, to the leading order in surface gravity $k_H \approx \frac{3\kappa^2(R_C-R_H)}{2R_H}$. Then,
\be
V(r^*)\approx\frac13 l(l+1)\frac{{k_H}^2}{(\kappa R_H)^2\cosh^2(k_H r^*)}+O({k_H}^3).
\ee
This is the P\"oschl-Teller potential, which makes the Schrodinger equation exactly solvable \cite{Ferrari:1984zz}. With the boundary condition $\Psi \sim e^{i\omega r^*}$ as $r^*\rightarrow -\infty$, we can get
\be
\Psi(r^*) = \left[\xi(1-\xi) \right]^{\frac{i\omega}{2 k_H}} {_2F_1} (1+\frac{i\omega}{k_H}+\beta,\frac{i\omega}{k_H}-\beta;1+\frac{i\omega}{k_H};\xi),
\ee
with $\xi\equiv\left(1+e^{-2 k_H r^*}\right)^{-1}$, $\beta\equiv i\sqrt{\frac{l(l+1)}{3(\kappa R_H)^2}-\frac14 }-\frac12$. As $r^*\rightarrow +\infty$ or $\xi\rightarrow 1$, the above solution has the asymptotic form
\be
\Psi(r^*) \sim C \left( e^{i\omega r^*} + R e^{-i\omega r^*} \right),
\ee
where $R$ and $C$ can be written in terms of Gamma functions \cite{Ferrari:1984zz}. The greybody factor is readily obtained as
\bea
&&\gamma_l(\omega)\equiv 1-|R|^2 \nonumber \\
&=& 1-\frac{|\sin(\pi\beta)\Gamma(-\beta+i\frac{\omega}{k_H})\Gamma(1+\beta+i\frac{\omega}{k_H})|^2}{\pi^2}.
\label{approx_extremal}
\eea
Note that this result doesn't hold for $l=0$, in which case the potential $V(r) \sim O({k_H}^3)$, thus the transmission coefficient will be much higher than the $l>0$ cases. On the other hand, for very large $l$ and low frequencies around the scale of $k_H$, $\gamma_l(\omega)$ vanishes, from the property of gamma functions,
\bea
\Gamma(-\beta+i\frac{\omega}{k_H})\Gamma(1+\beta+i\frac{\omega}{k_H})&\approx&\Gamma(-\beta)\Gamma(1+\beta)\nonumber\\
&=&-\frac{\pi}{\sin(\pi\beta)}.
\eea

At high energy and $l \gtrsim 10$, then the arguments in all functions in Eq. (\ref{approx_extremal}) are practionally imaginary, making further simplification possible. Using the asymptotic approximation
\be
\Gamma(ix)\approx e^{-ix}(ix)^{ix}\left(\frac{2\pi}{ix} \right)^{1/2}(1+O(1/x)),
\ee
we can put $\gamma_l(\omega)$ in a cleaner form for $\omega$ not close to $\frac{l k_H}{\sqrt{3}\kappa R_H}$, so that arguments of the gamma functions in Eq. (\ref{approx_extremal}) are both imaginarily large. Therefore, keeping only the leading order term,
\bea
&&\gamma_l(\omega) \nonumber\\
&\approx& 1- \frac{4 e^{-\pi\left(\left|\frac{\omega}{k_H}+\frac{l}{\sqrt{3}\kappa R_H}\right|+\left|\frac{\omega}{k_H}-\frac{l}{\sqrt{3}\kappa R_H}\right|\right)}}{\left|(\frac{\omega}{k_H}+\frac{l}{\sqrt{3}\kappa R_H})(\frac{\omega}{k_H}-\frac{l}{\sqrt{3}\kappa R_H})\right|} \sinh^2\left(\frac{\pi l}{\sqrt{3}\kappa R_H}\right) \nonumber \\
&=& 1-\frac{4 e^{-2\pi \max\left(\frac{\omega}{k_H},\frac{l}{\sqrt{3}\kappa R_H}\right)}}{\left|\left(\frac{\omega}{k_H}\right)^2-\frac{l^2}{3(\kappa R_H)^2}\right|} \sinh^2\left(\frac{\pi l}{\sqrt{3}\kappa R_H}\right).
\eea
Thus for $\omega\gg\frac{l k_H}{\sqrt{3}\kappa R_H}$,
\be
\gamma_l(\omega)\approx 1-\left(\frac{k_H}{\omega} \right)^2 e^{-2\pi\omega/k_H},
\ee
which is exponentially approaching 1 as $\omega$ increases. And for $k_H\ll\omega\ll\frac{l k_H}{\sqrt{3}\kappa R_H}$,
\be
\gamma_l(\omega)\approx 1-\frac{3(\kappa R_H)^2}{l^2},
\ee
which is independent of $\omega$, but approaching 1 quadratically as $l$ increases.

For $l$ from 1 to 8, our analytical expression will be shown to be perfectly consistent with the numerical results in figure (\ref{extremal}).

\section{Coupling with background fields}
\label{cbg}
As we have demonstrated in previous sections, without the coupling of the test scalar field to the background {St\"uckelberg} fields, calculations effectively reduce to the Schwarzschild-dS black hole case. In this section, we take the coupling into consideration.
Massive gravitons can be decomposed into tensor modes, as well as vector and scalar modes $A^{\mu}$ and $\pi$, which turn out to be the non-unitary parts of the background {St\"uckelberg} fields \cite{Berezhiani:2011mt}, i.e. $x^{\mu}-\phi^{\mu}=(m_g A^{\mu}+\partial^{\mu}\pi)/\Lambda^3$. In the Schwarzschild coordinates, the vector modes were found to be
\bea
A^i&=&0 , \\
A^0&=& -\frac{M_{pl}m_g}{\kappa_0}f(r),
\eea
and the scalar mode $\pi$ depends on both $r$ and $t$. Here the free dimensionless parameter $\kappa_0$ is an integration constant when solving the Einstein equation to obtain the Schwarzschild-dS metric \cite{Berezhiani:2011mt}, and
\be
f(r)= \pm \int dr\frac{\sqrt{\frac{r_g}r+\frac2{3\alpha}m^2 r^2}}{1-\frac{r_g}r-\frac2{3\alpha}m^2 r^2},
\ee
which diverges at both horizons.

The $\pi$ mode depends on the coordinate time \cite{Berezhiani:2011mt}, so coupling with it will destroy our assumed solution form in (\ref{seperate}). So for simplicity, we focus on the interaction with the vector modes $A^{\mu}$, and further require invariance under the Lorentz and discrete $\varphi\rightarrow -\varphi$ transformations, and renormalizability. This leaves only $\sqrt{-g}A^2 \varphi^2$ and
$\sqrt{-g}\varphi\partial_{\mu}\varphi A^{\mu}$ couplings in the lagrangian density. However,
\be
\sqrt{-g}\varphi\partial_{\mu}\varphi A^{\mu}=\frac12 \partial_{\mu}(\sqrt{-g}\varphi^2 A^{\mu})-\frac12\varphi^2 \partial_{\mu}(\sqrt{-g}A^{\mu}),
\ee
where the first term on the right-hand side is a total derivative thus not contributing to the equations of motion of $\varphi$, and the second term vanishes because $g_{\mu\nu}$ and the only nonzero component of the vector mode $A^0$ do not depend on the coordinate time. Therefore, from now on, we consider only the interaction term
\be
\mathcal{L}_{coupling}= \frac12\sqrt{-g}\left(\lambda A^2 \varphi^2\right).
\ee
Here $\lambda$ is a dimensionless coupling constant. This form of coupling is the only possible renormalizable form consistent with the symmetries in the model. It would break any eventual gauge invariance for the field $A^{\mu}$, since it will act as an effective mass for the field $A^{\mu}$, but in massive gravity this field is massive anyway.

\begin{figure}[h]
  \centering
\includegraphics[height=0.27\textwidth,angle=0]{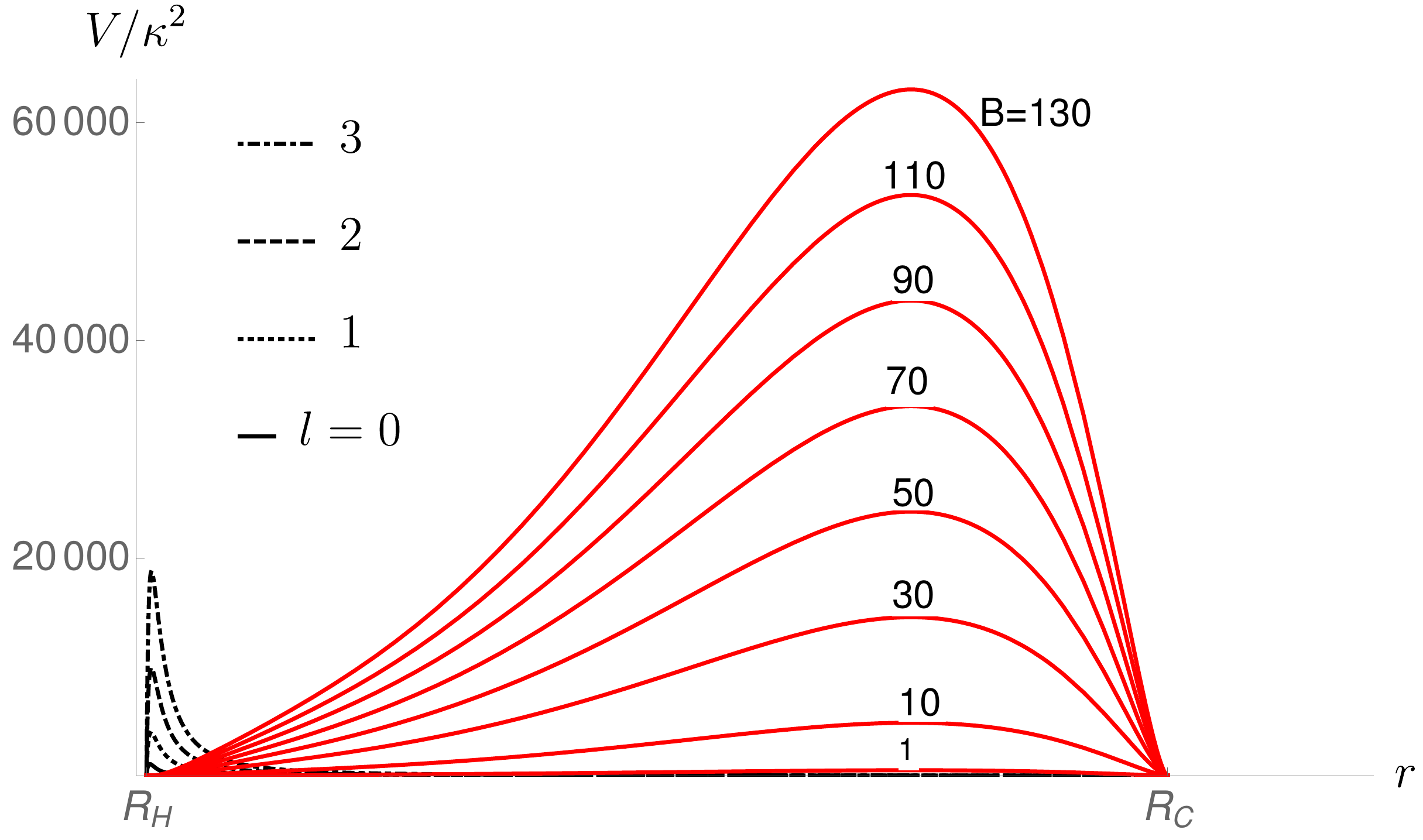}
\caption{Effective potential for $\kappa R_H=0.01$ and different values of $l$ and coupling strength $B$. Two terms contributes, one due to the black hole geometry, i.e. Eq. (\ref{V(r)}) (left crests, in black), and the other due to coupling with background fields, i.e. Eq. (\ref{V_coupling}) (right crests, in red).}
\label{V_small}
\end{figure}

This coupling term induces an additional potential term in Eq. (\ref{V(r)}),
\bea
\label{V_coupling}
V_{coupling}(r) &=& -\lambda A^2(r) F(r) \\
&=& \lambda\left(\frac{M_{pl}m_g}{\kappa_0}\right)^2 \left[F(r)\int dr \frac{\sqrt{1-F(r)}}{F(r)} \right]^2.\nonumber
\eea
The integral term has logarithmic divergences at both horizons. However it is multiplied by the $F(r)$ term in front of it which fixes this divergence, thus enabling us to use the techniques in section II to compute the greybody factors in this case. The forms of the effective potentials are shown for the small black holes in figure (\ref{V_small}).
 Here, the contributions from black hole geometry, i.e. Eq. (\ref{V(r)}), are seen as the spikes on the left side of the figure, while the contributions from the coupling with massive gravitons are shown as the spikes on the right part.

\begin{figure*}
\def\tabularxcolumn #1{m{#1}}
\begin{tabularx}{\linewidth}{@{}cXX@{}}
\begin{tabular}{ccc}
\captionsetup[subfigure]{labelformat=empty}
\subfloat[$l=0$]{\includegraphics[width=0.30\textwidth,angle=0]{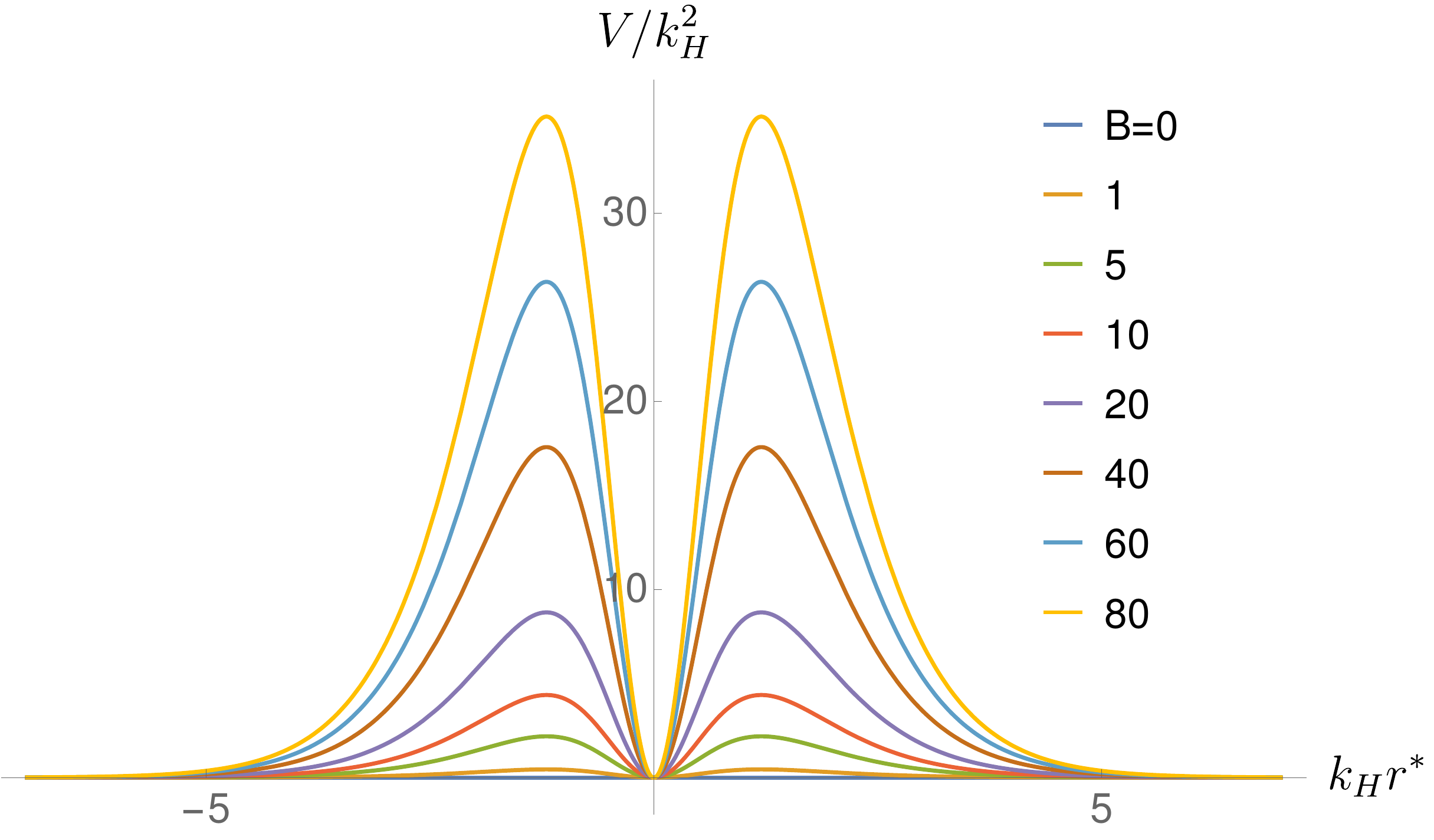}}
&\captionsetup[subfigure]{labelformat=empty}
 \subfloat[$l=1$]{\includegraphics[width=0.30\textwidth,angle=0]{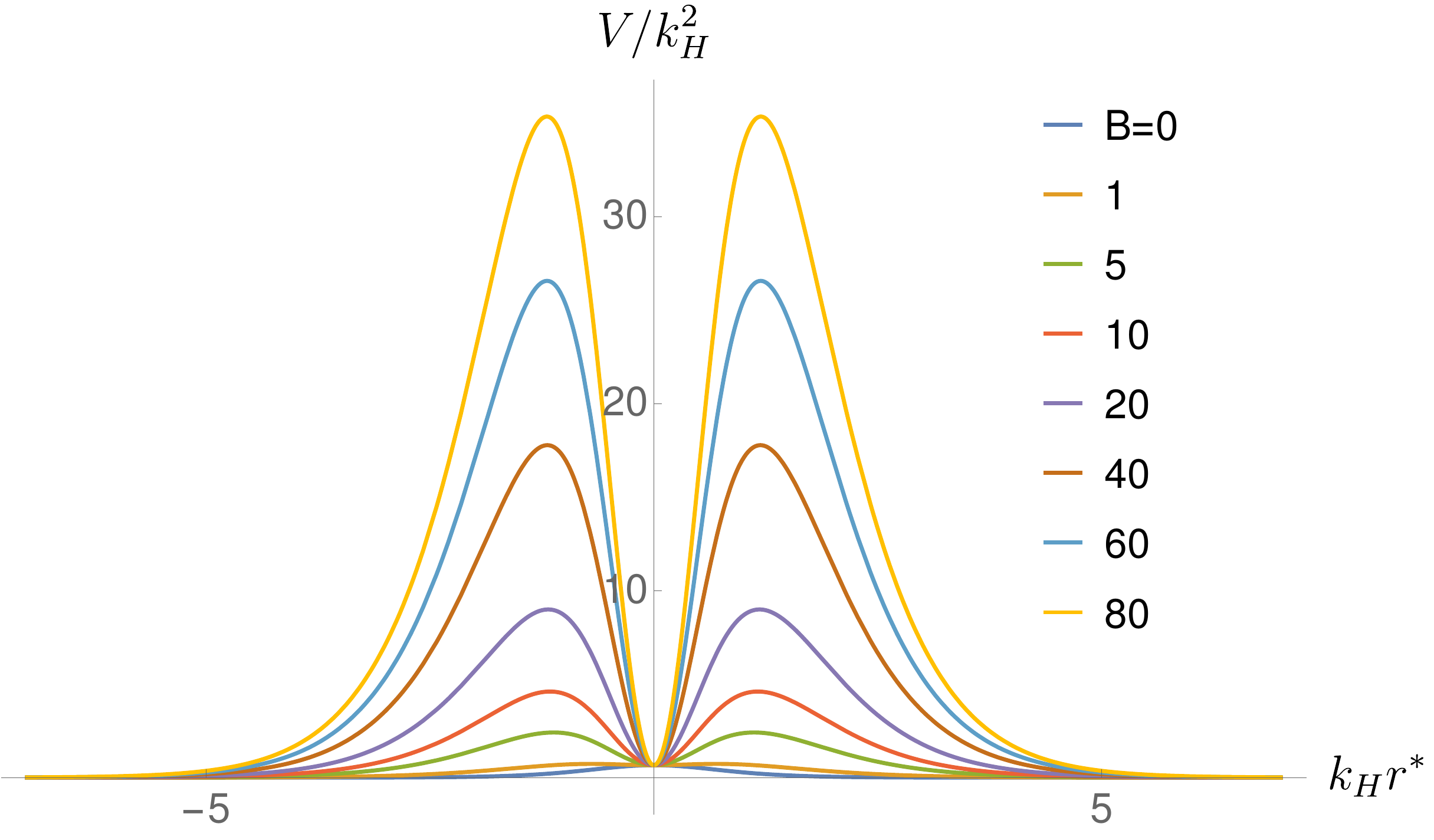}}
&\captionsetup[subfigure]{labelformat=empty}
\subfloat[$l=2$]{\includegraphics[width=0.30\textwidth,angle=0]{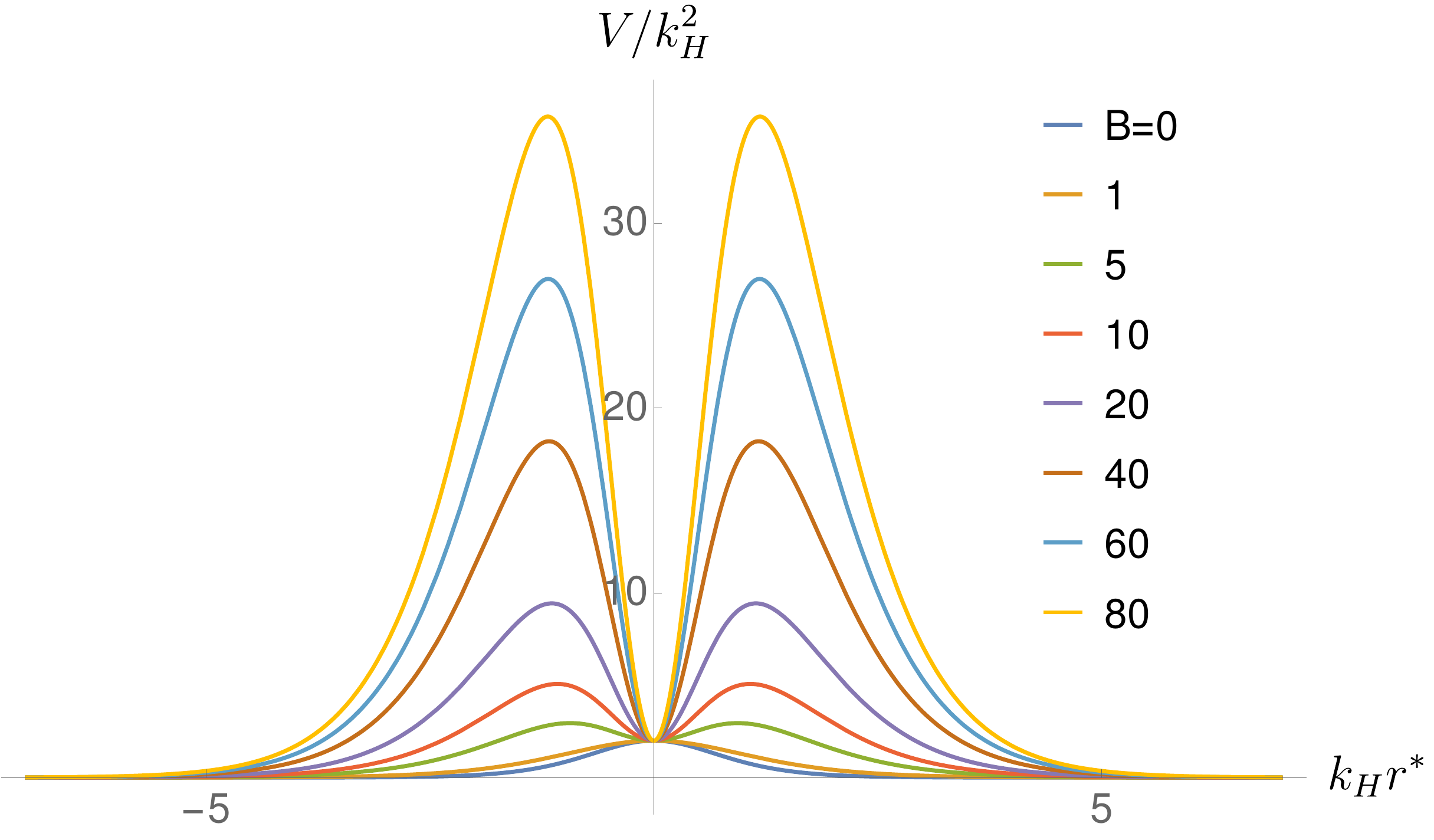}} \\
\captionsetup[subfigure]{labelformat=empty}
 \subfloat[$l=3$]{\includegraphics[width=0.30\textwidth,angle=0]{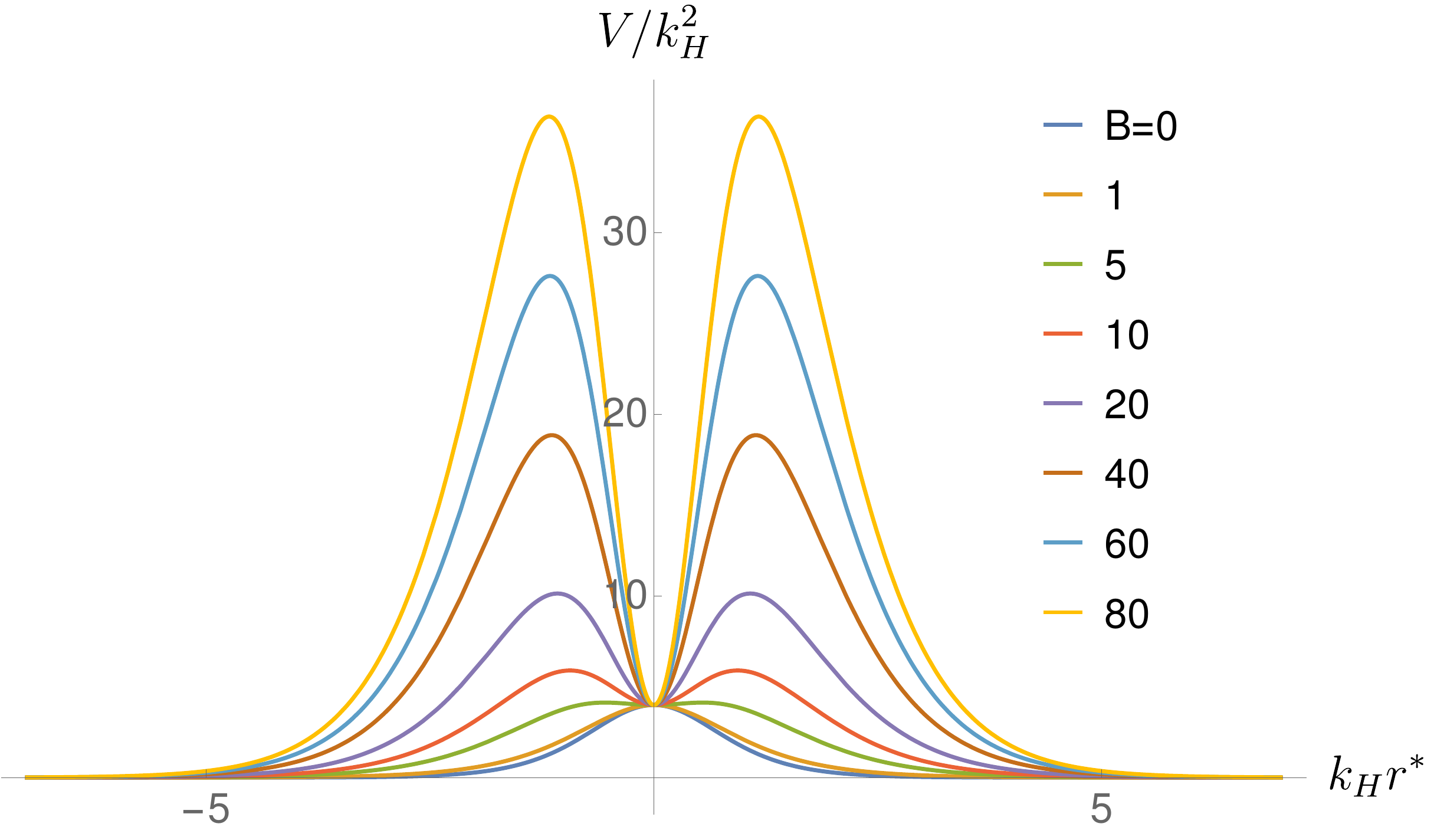}}
&\captionsetup[subfigure]{labelformat=empty}
 \subfloat[$l=4$]{\includegraphics[width=0.30\textwidth,angle=0]{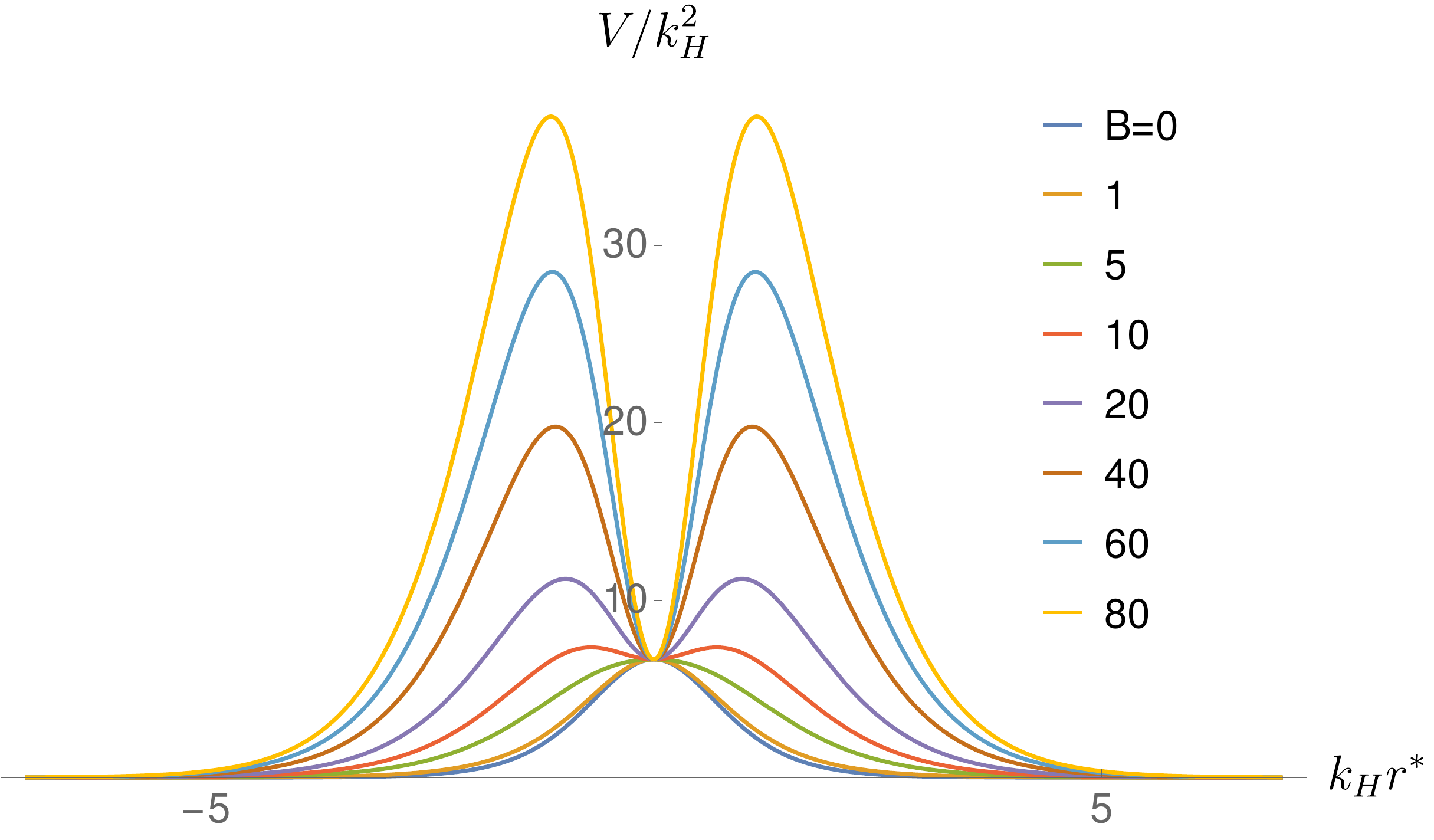}}
&\captionsetup[subfigure]{labelformat=empty}
 \subfloat[$l=5$]{\includegraphics[width=0.30\textwidth,angle=0]{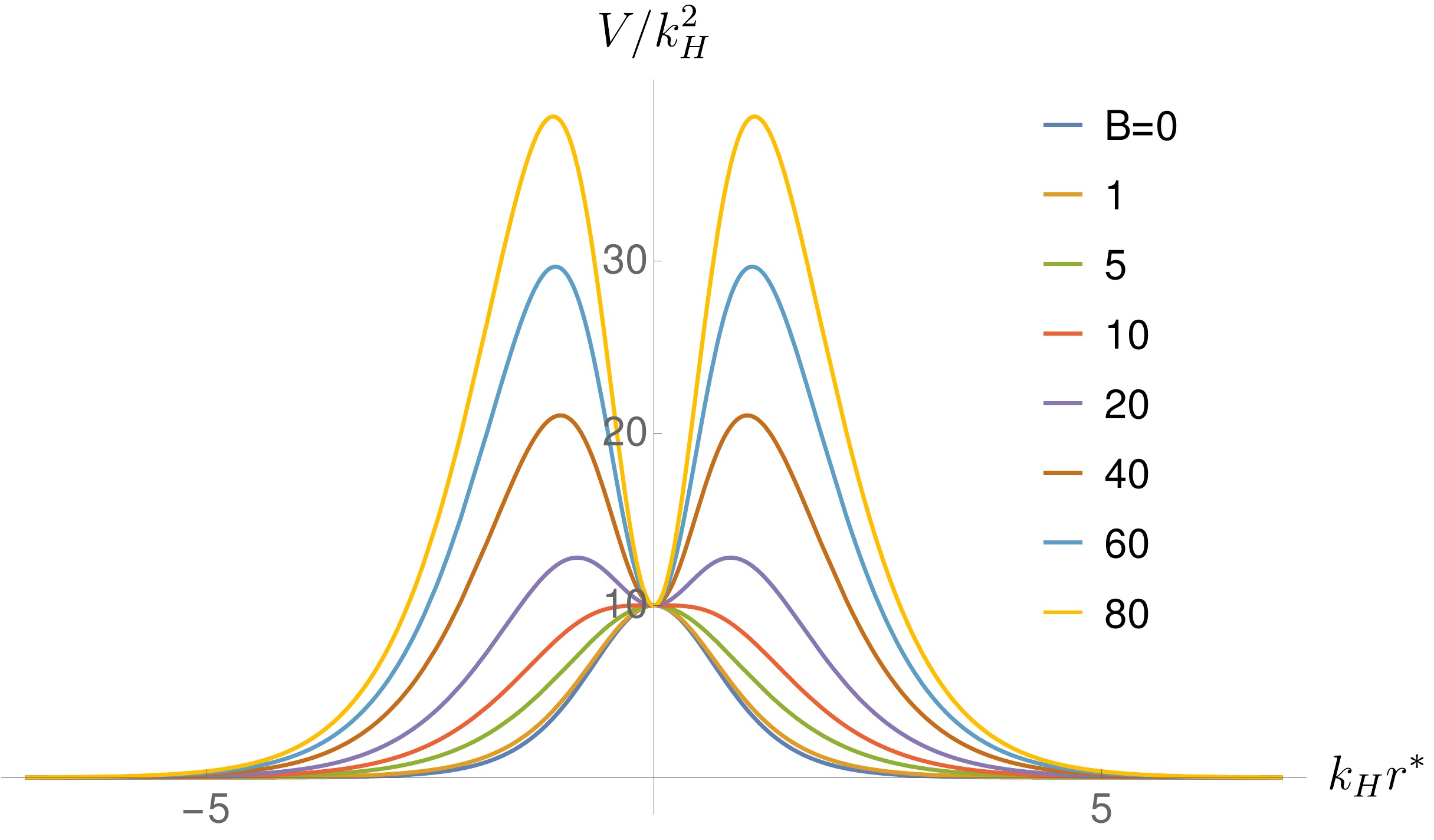}} \\
\captionsetup[subfigure]{labelformat=empty}
 \subfloat[$l=6$]{\includegraphics[width=0.30\textwidth,angle=0]{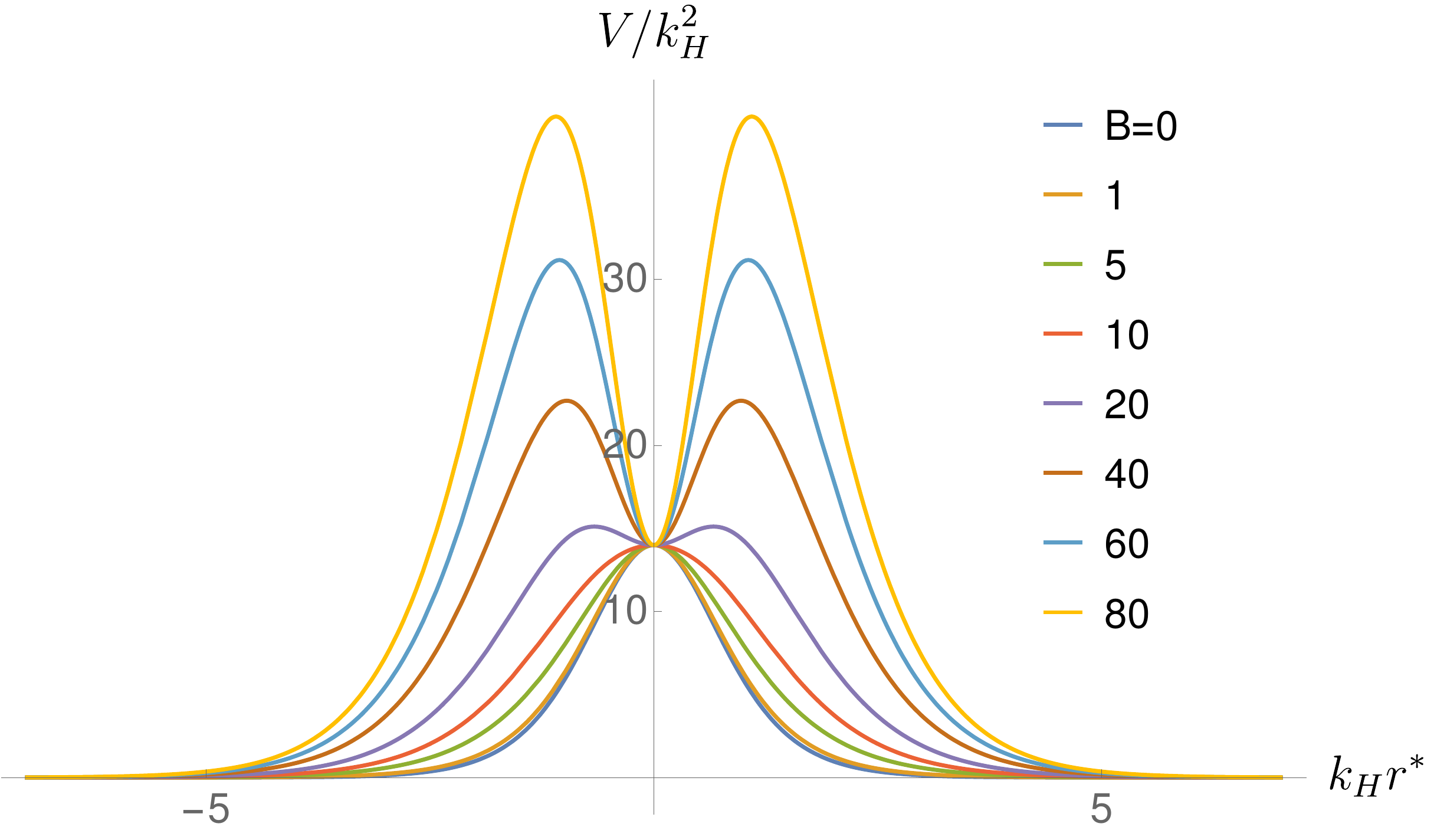}}
&\captionsetup[subfigure]{labelformat=empty}
 \subfloat[$l=7$]{\includegraphics[width=0.30\textwidth,angle=0]{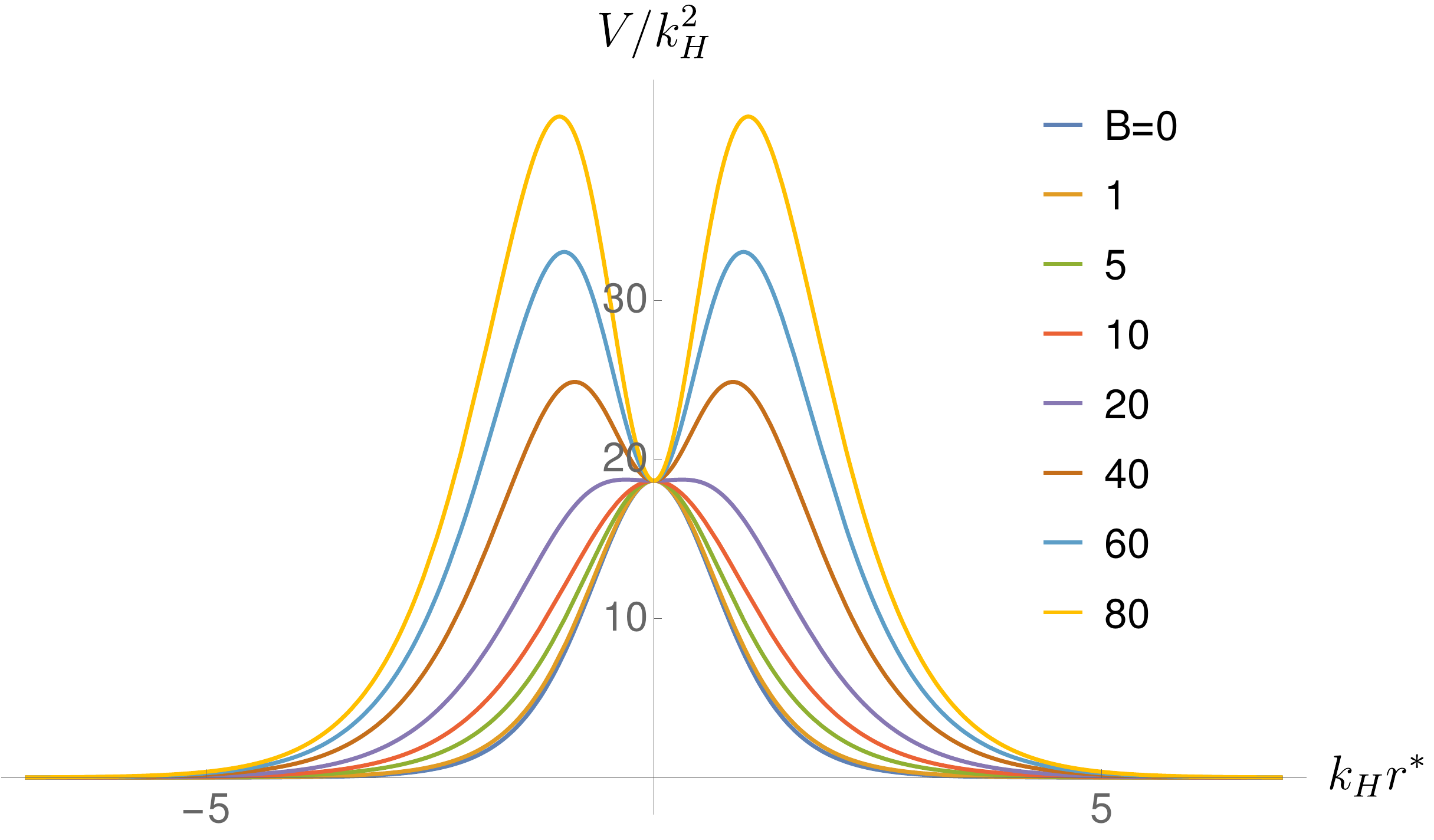}}
&\captionsetup[subfigure]{labelformat=empty}
 \subfloat[$l=8$]{\includegraphics[width=0.30\textwidth,angle=0]{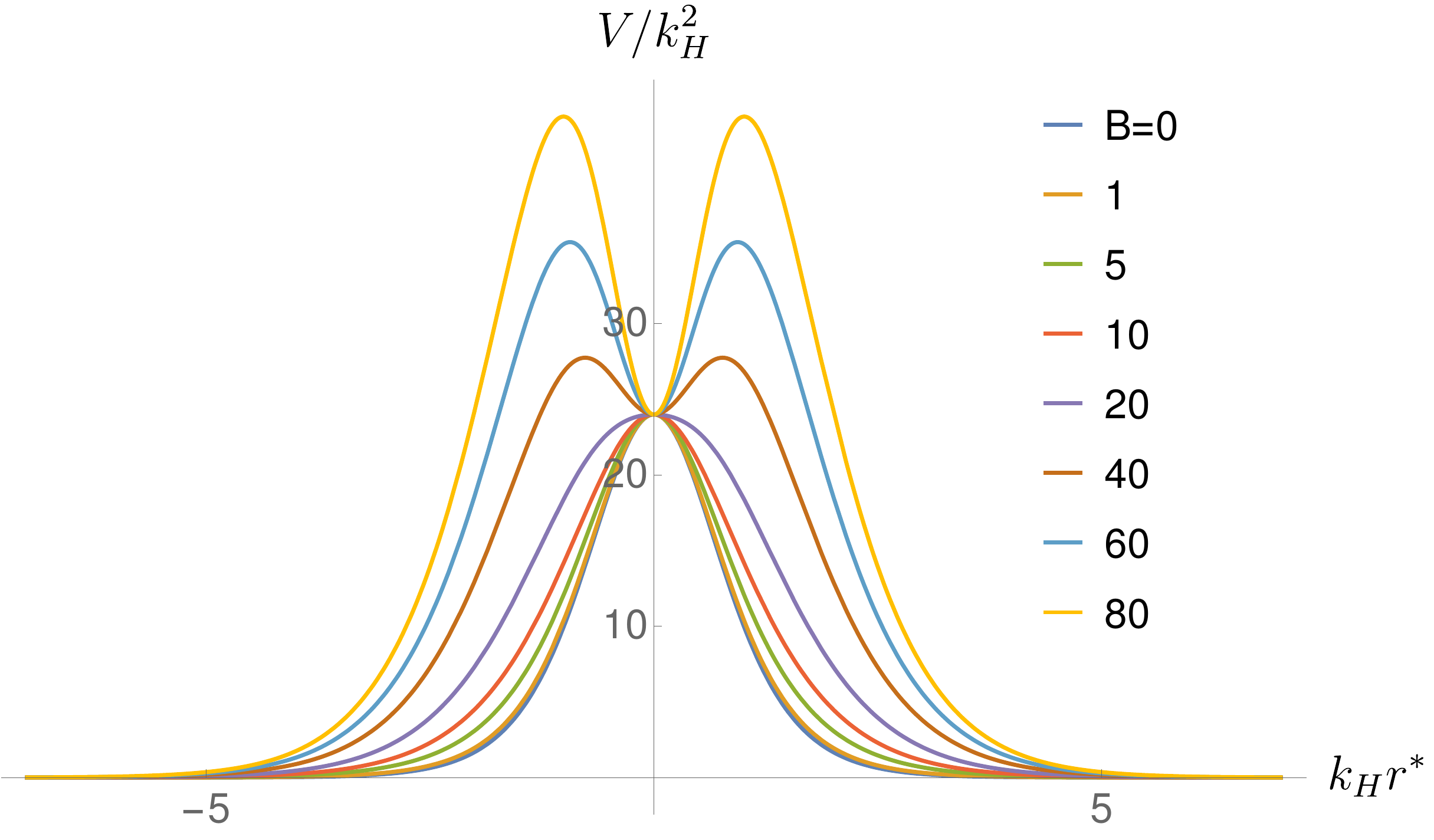}}
\end{tabular}
\end{tabularx}
\caption{Effective potentials for near-extremal black holes, with various coupling strength between test scalars and the background vector fields, and $l$ up to 8. Different couplings are differentiated by different colors.}
\label{V_extremal}
\end{figure*}

The total effective potential for near-extremal black holes is shown in Fig. (\ref{V_extremal}) for $l$ up to 8. For $B=0$, the potential is P\"oschl-Teller, and becomes effectively wider and wider as $B$ increases.

\begin{figure*}
\def\tabularxcolumn #1{m{#1}}
\begin{tabularx}{\linewidth}{@{}cXX@{}}
\begin{tabular}{cc}
\captionsetup[subfigure]{labelformat=empty}
\subfloat[$l=0$]{\includegraphics[width=0.45\textwidth,angle=0]{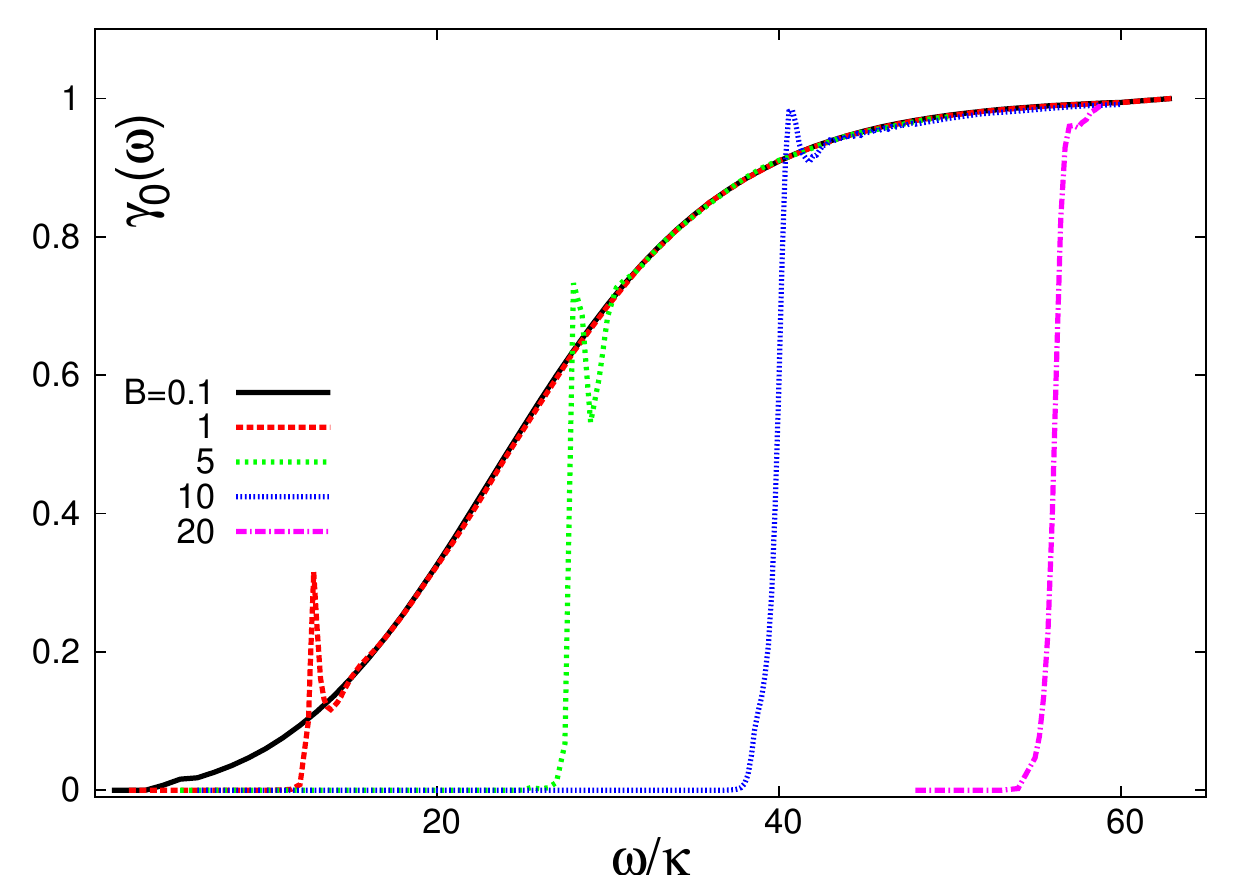}}
&\captionsetup[subfigure]{labelformat=empty}
 \subfloat[$l=1$]{\includegraphics[width=0.45\textwidth,angle=0]{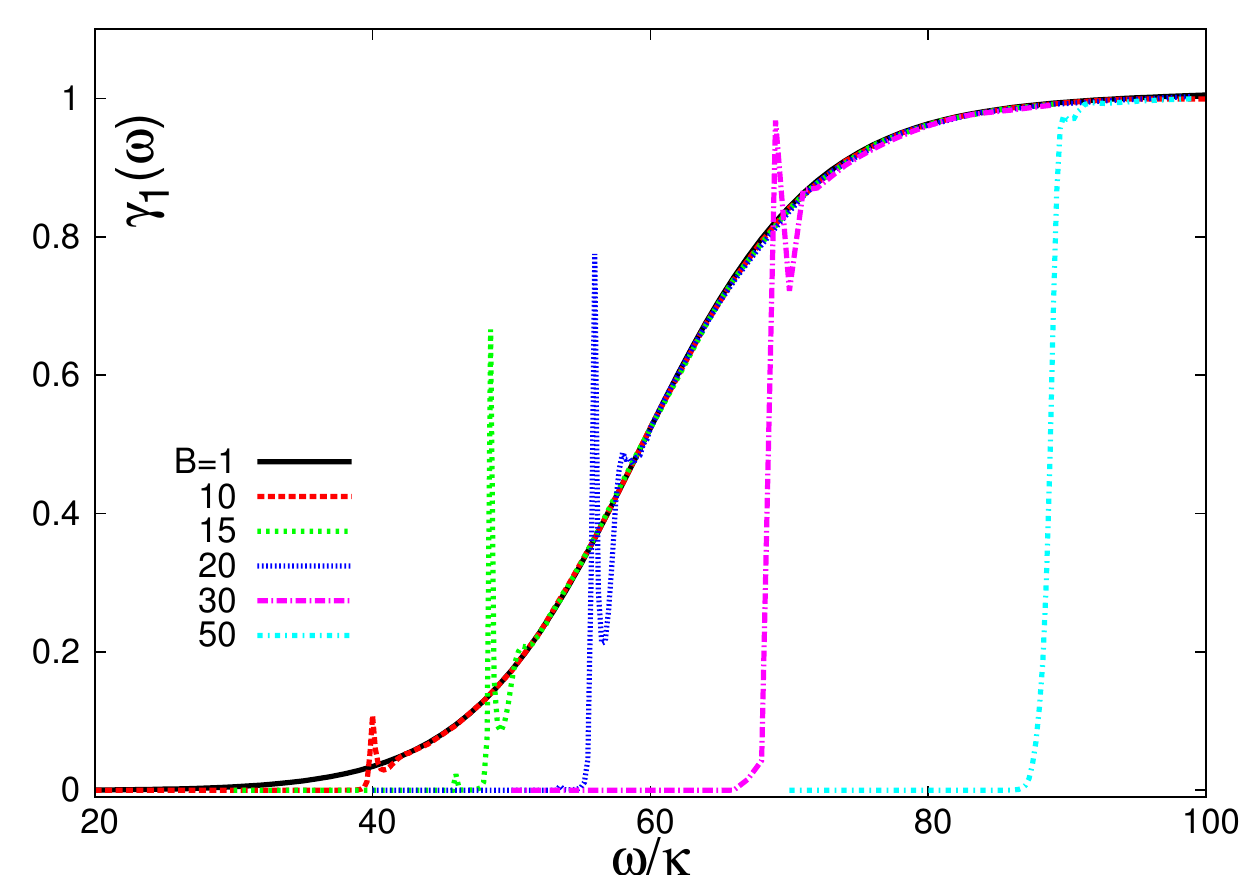}} \\
\captionsetup[subfigure]{labelformat=empty}
\subfloat[$l=2$]{\includegraphics[width=0.45\textwidth,angle=0]{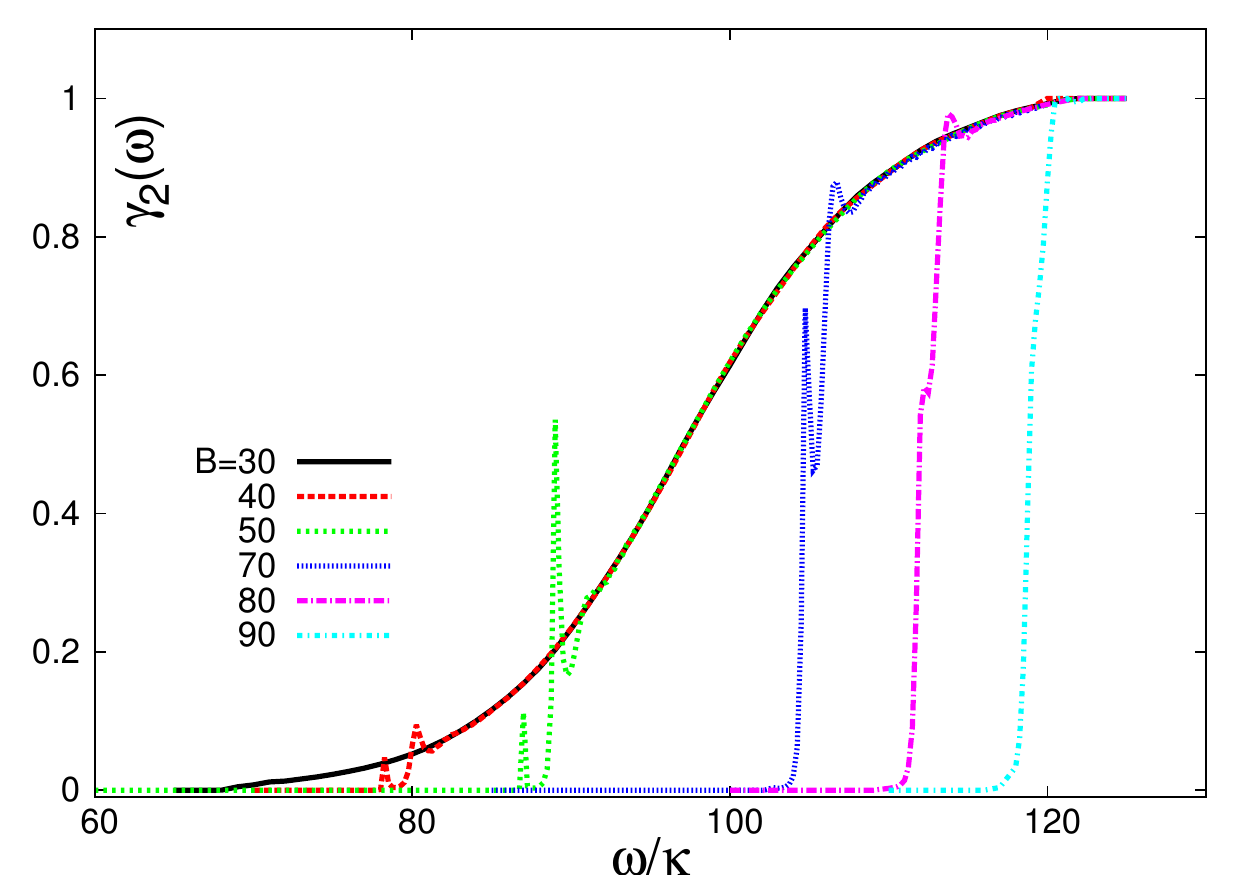}}
&\captionsetup[subfigure]{labelformat=empty}
 \subfloat[$l=3$]{\includegraphics[width=0.45\textwidth,angle=0]{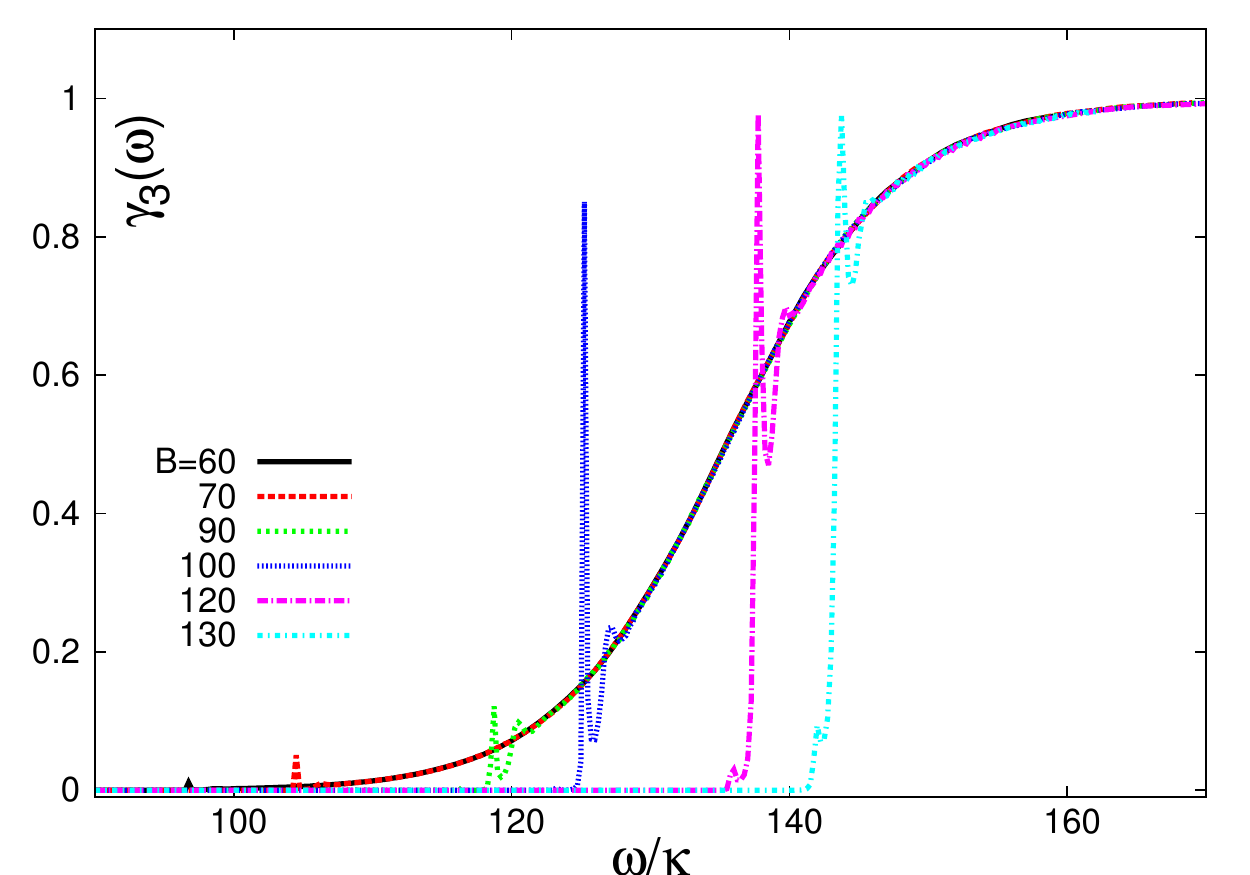}}
\end{tabular}
\end{tabularx}
\caption{Greybody factors for small black holes $\kappa R_H=0.01$, with coupling between test scalars and background vector modes. The spikes in the greybody factors are due to resonances in transmission rate which in principle increase the transmission and allow more
particles to penetrate the barrier.}
\label{small}
\end{figure*}

\begin{figure*}
\def\tabularxcolumn #1{m{#1}}
\begin{tabularx}{\linewidth}{@{}cXX@{}}
\begin{tabular}{ccc}
\captionsetup[subfigure]{labelformat=empty}
\subfloat[$l=0$]{\includegraphics[width=0.30\textwidth,angle=0]{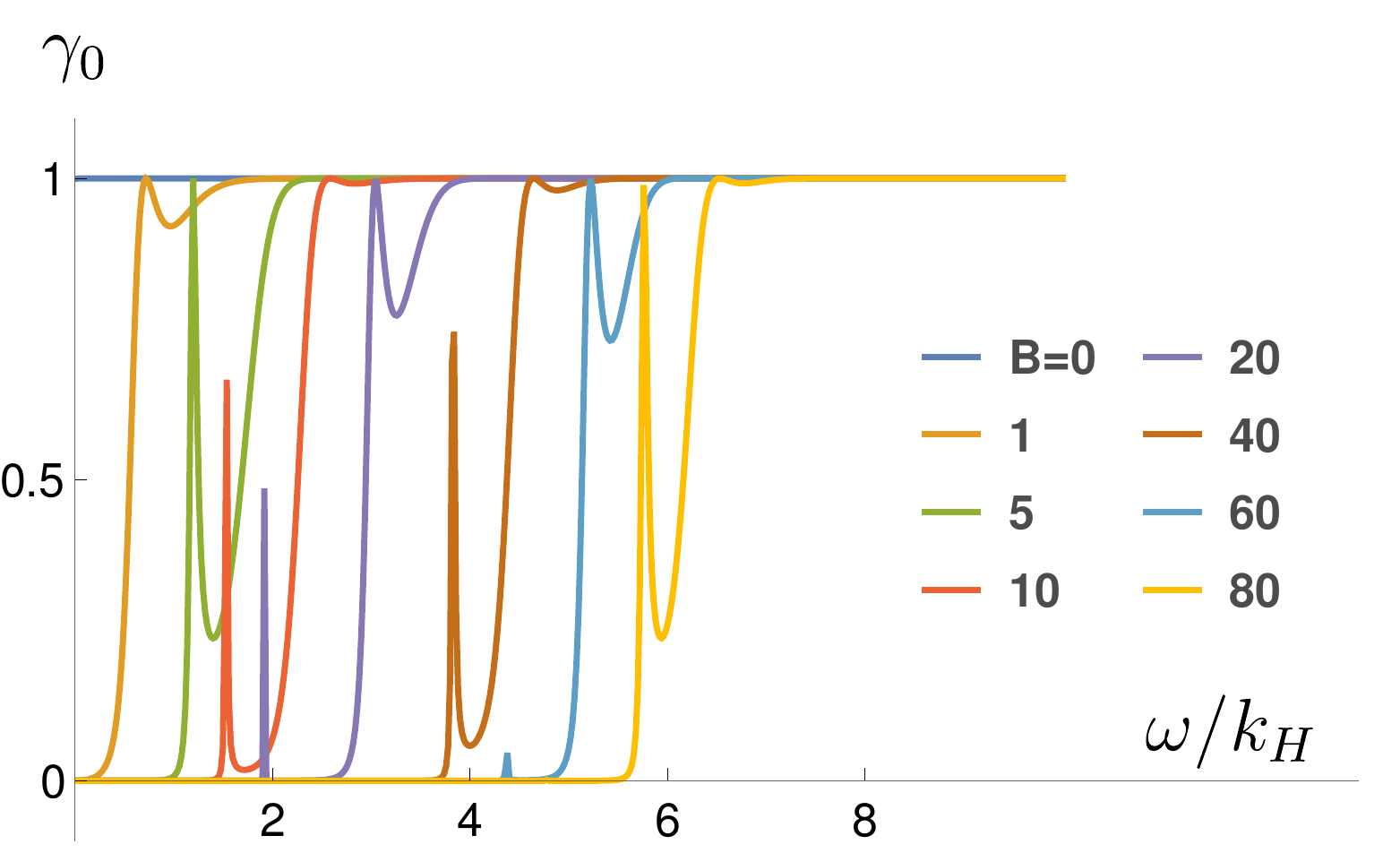}}
&\captionsetup[subfigure]{labelformat=empty}
 \subfloat[$l=1$]{\includegraphics[width=0.30\textwidth,angle=0]{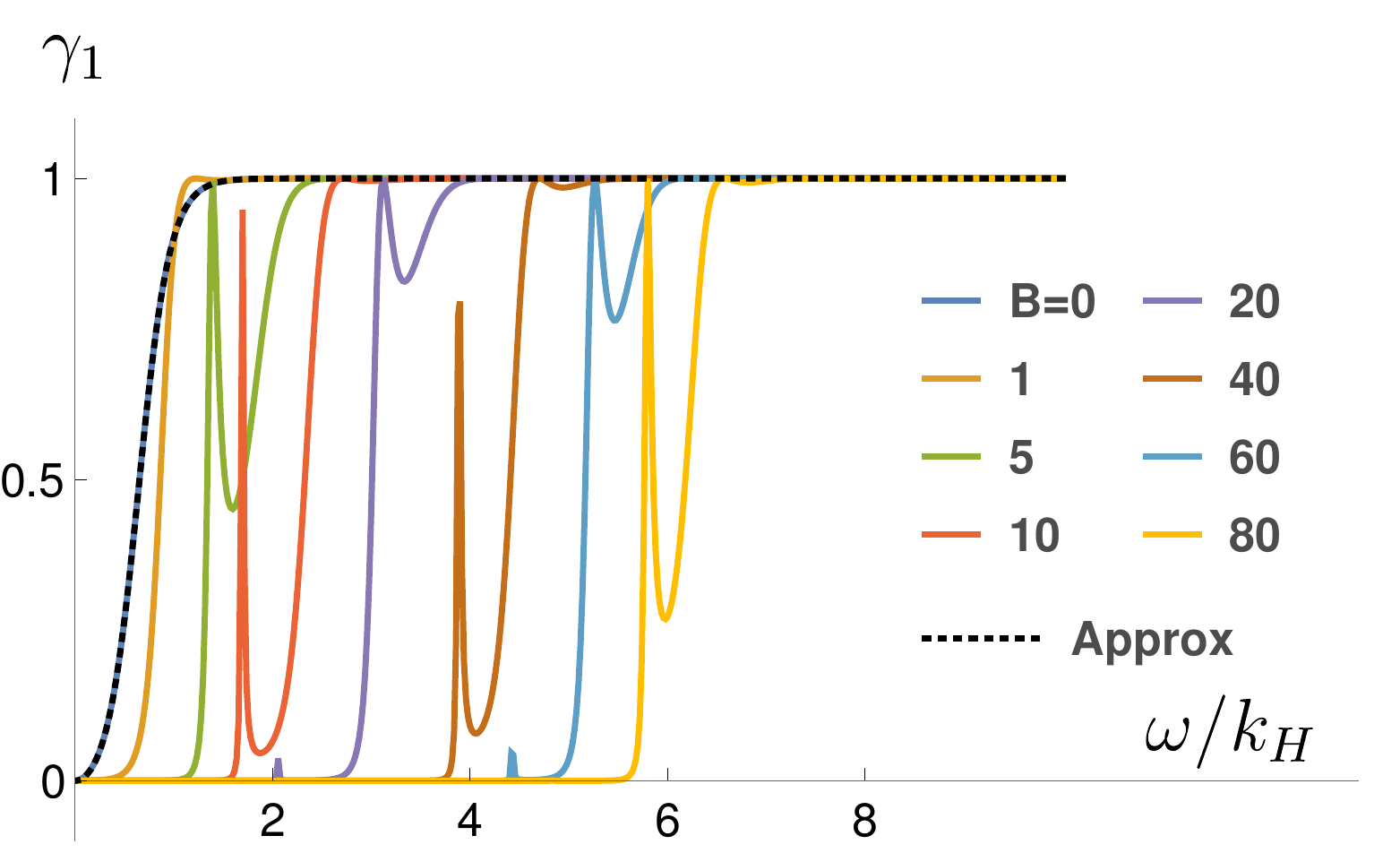}}
&\captionsetup[subfigure]{labelformat=empty}
\subfloat[$l=2$]{\includegraphics[width=0.30\textwidth,angle=0]{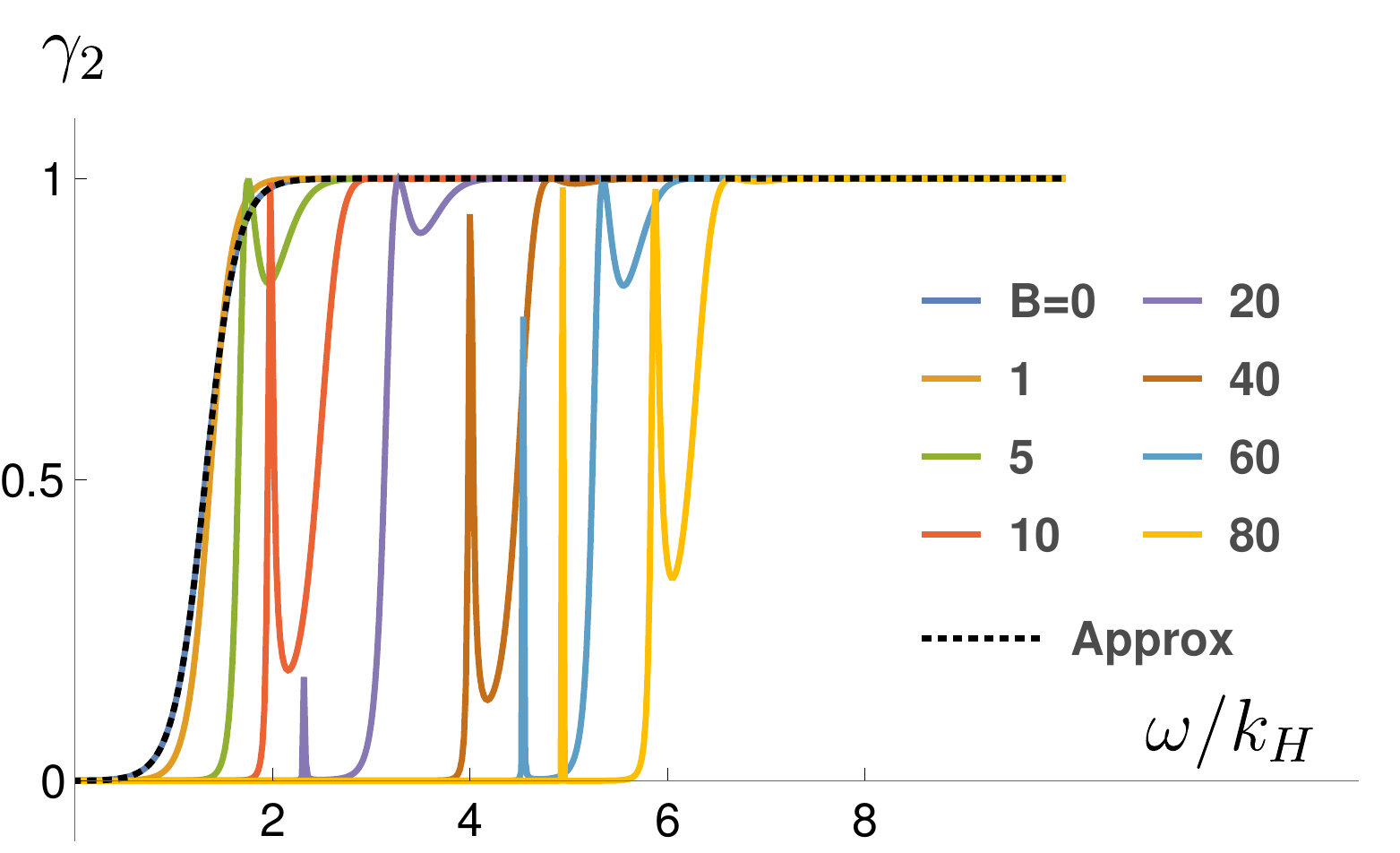}} \\
\captionsetup[subfigure]{labelformat=empty}
\subfloat[$l=3$]{\includegraphics[width=0.30\textwidth,angle=0]{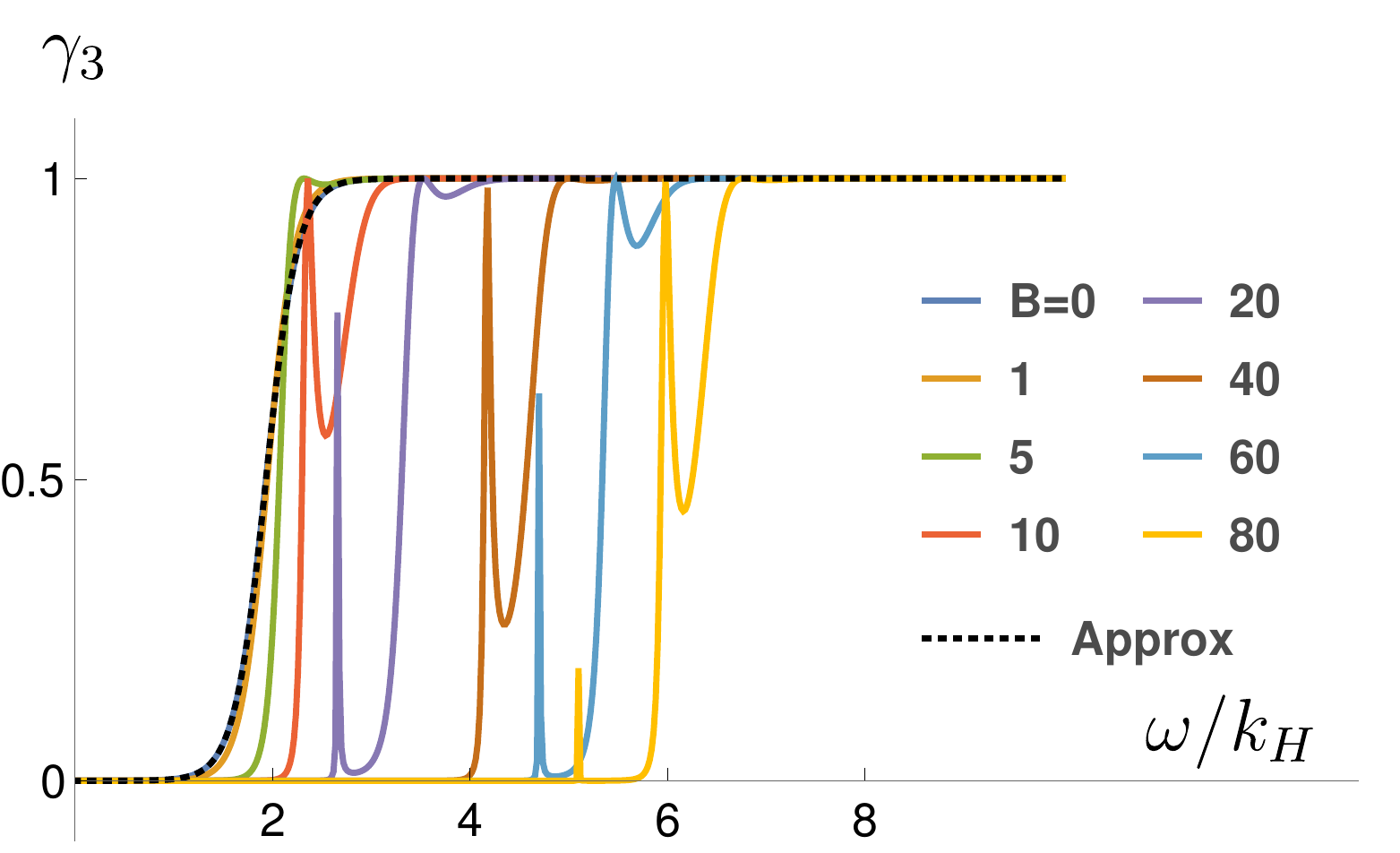}}
&\captionsetup[subfigure]{labelformat=empty}
 \subfloat[$l=4$]{\includegraphics[width=0.30\textwidth,angle=0]{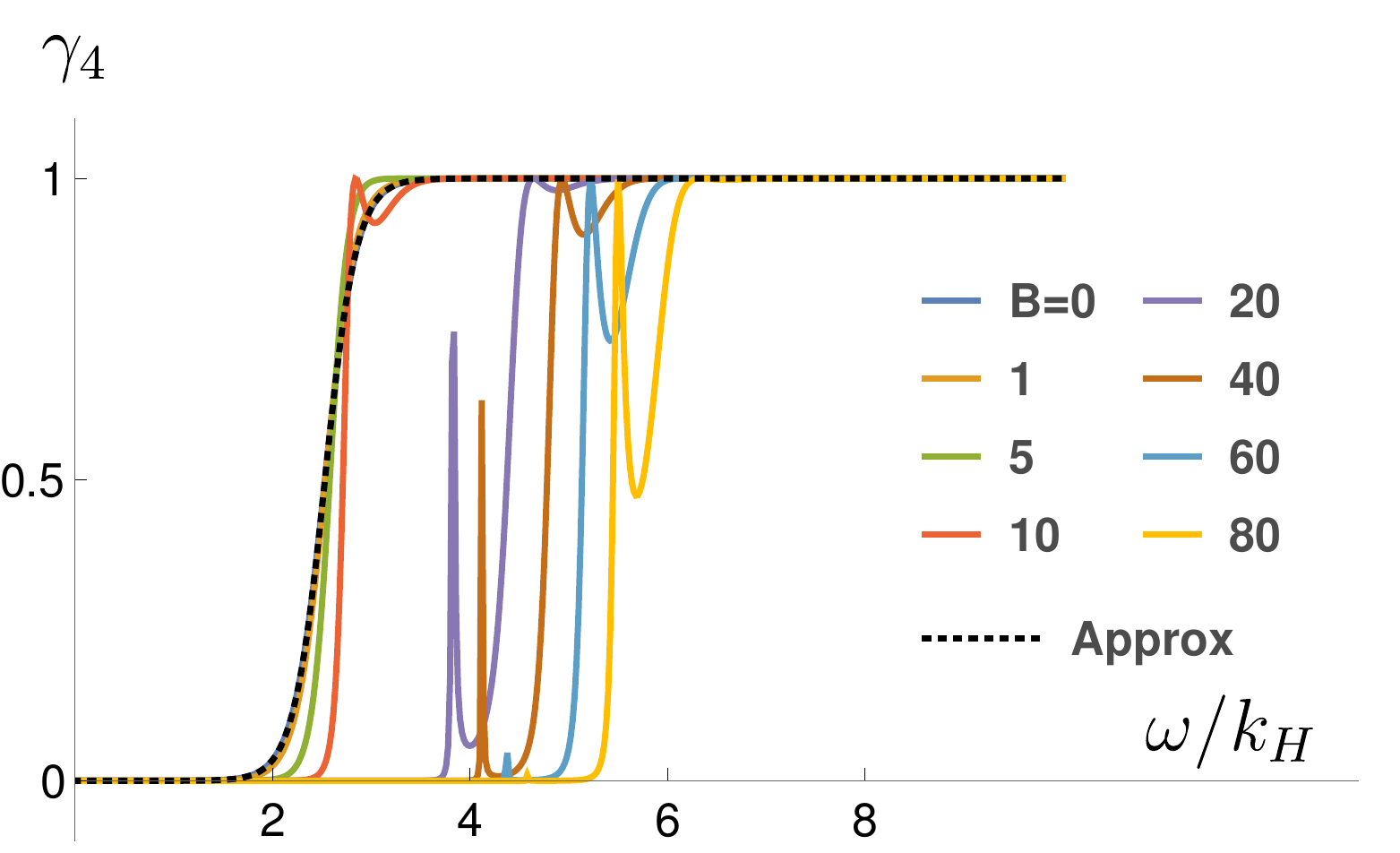}}
&\captionsetup[subfigure]{labelformat=empty}
\subfloat[$l=5$]{\includegraphics[width=0.30\textwidth,angle=0]{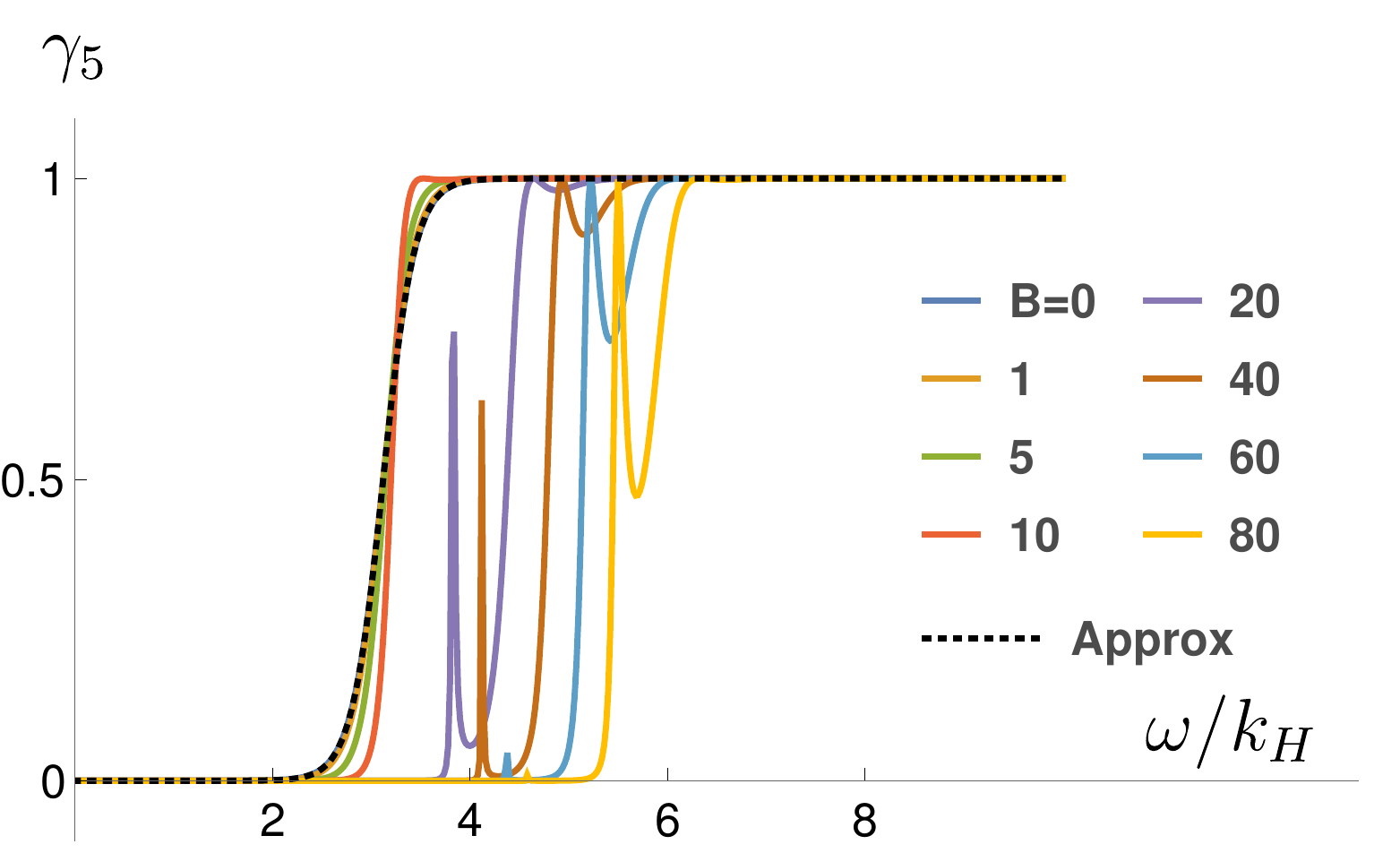}} \\
\captionsetup[subfigure]{labelformat=empty}
\subfloat[$l=6$]{\includegraphics[width=0.30\textwidth,angle=0]{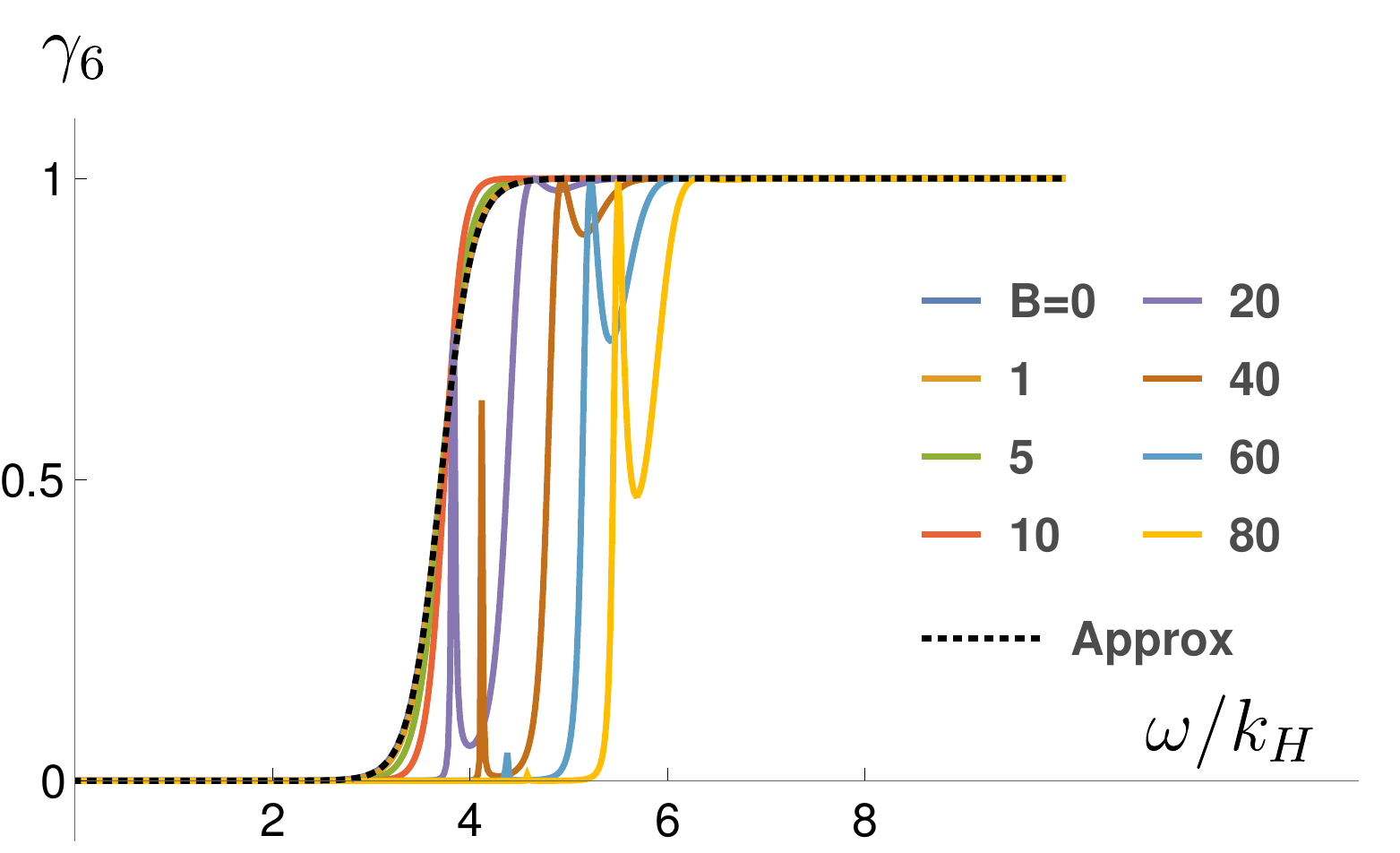}}
&\captionsetup[subfigure]{labelformat=empty}
 \subfloat[$l=7$]{\includegraphics[width=0.30\textwidth,angle=0]{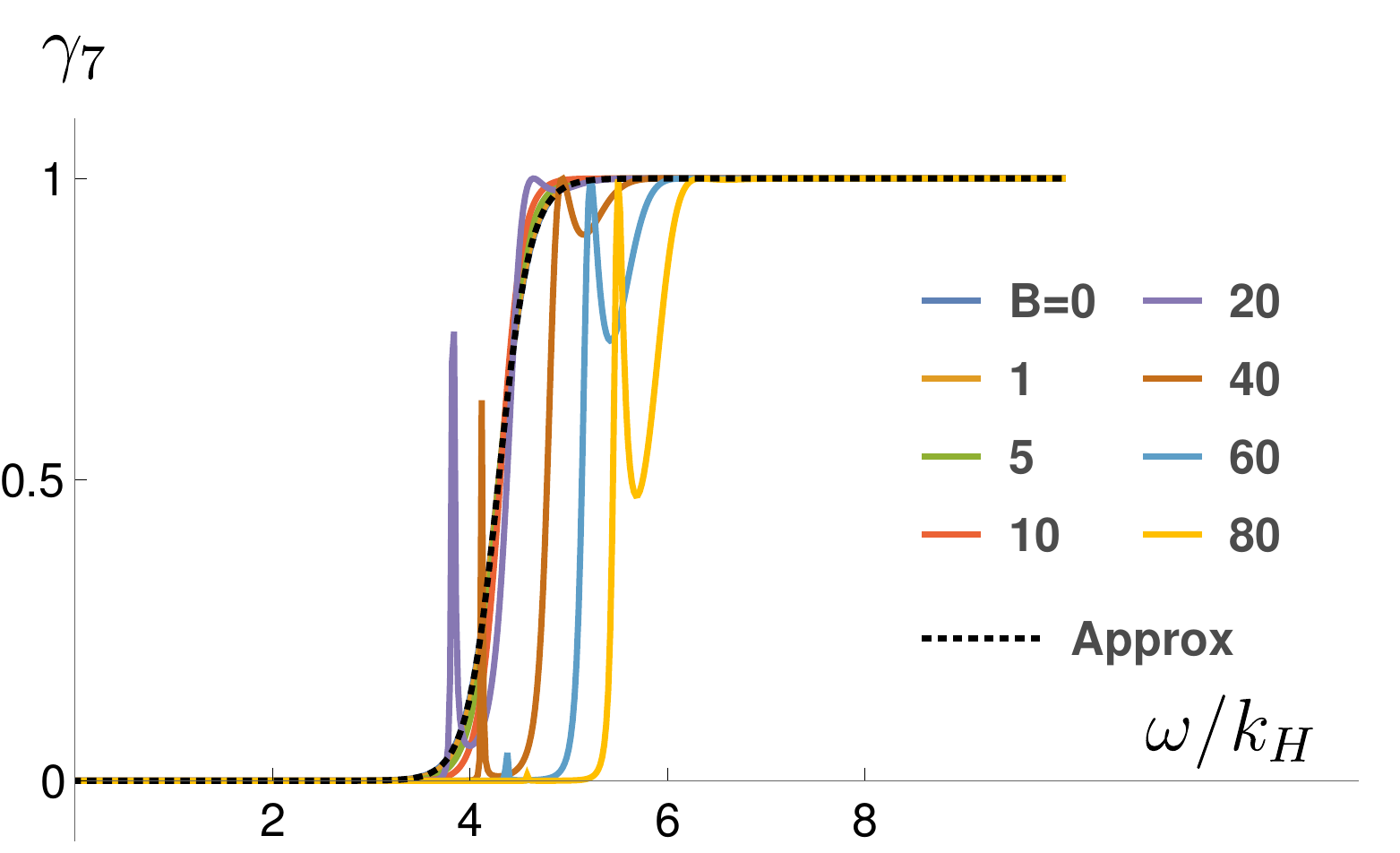}}
&\captionsetup[subfigure]{labelformat=empty}
\subfloat[$l=8$]{\includegraphics[width=0.30\textwidth,angle=0]{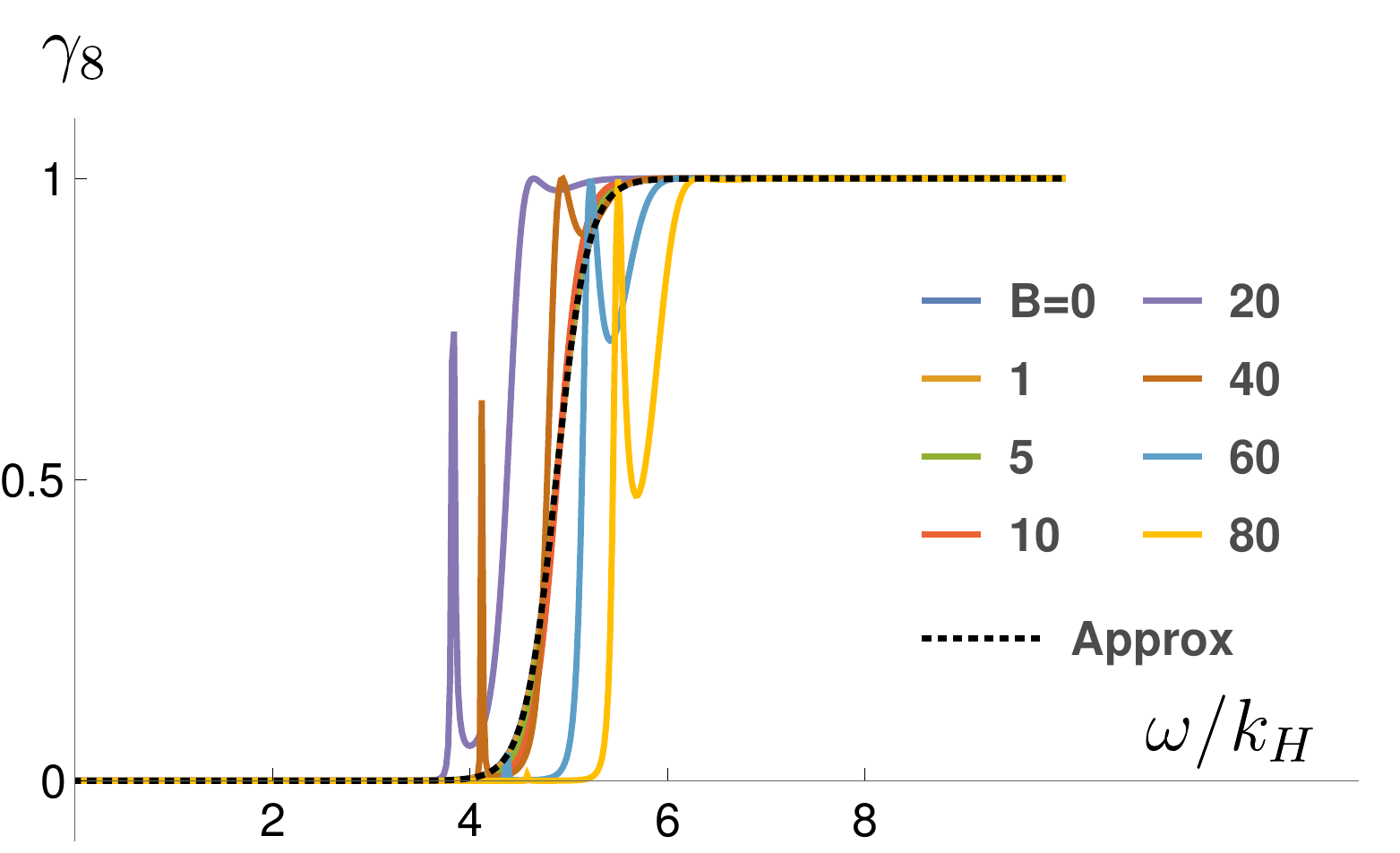}}
\end{tabular}
\end{tabularx}
\caption{Greybody factors for near-extremal black holes, with various coupling strength between test scalars and background vector modes. The black dashed lines are the analytical approximation of Eq. (\ref{approx_extremal}) for zero coupling,
which can be seen to coincide with our numerical calculations. Different couplings are differentiated by different colors.}
\label{extremal}
\end{figure*}

The same step sizes are used as in section III \footnote{A stepsize of $10^{-4}$ is also sufficient for the numerical calculation for $l$ up to 8 and $B$ up to 80, in the near-extremal case.}. The results for greybody factors are shown for both small and near-extremal black holes in figures (\ref{small}) and (\ref{extremal}).

It turns out that effect of the coupling can be measured by a parameter $B\equiv \frac{\lambda}3 \left(\frac{M_{pl}m_g}{\kappa_0 \kappa k_H} \right)^2 = \frac{\lambda \alpha}2 \left(\frac{M_{pl}}{\kappa_0 k_H} \right)^2$, which is independent of the graviton mass. Note that $\alpha$ and $\kappa_0$ are dimensionless parameters in massive gravity. In the small black hole case, $\gamma_l$ is moved to higher frequencies
as $B$ gets larger due to higher potential barrier, similar to increasing $l$. This argument also holds for near-extremal black holes. Besides, nonzero coupling will also make $\gamma_l$ show additional peaks and dips at some frequencies, for relatively large values of $B$ and smaller values of $l$, for both small and near-extremal black holes.

To explain the appearance of the visible peaks (and associated dips) in the near-extremal case, we write explicitly the form of the effective potential appearing in equation (\ref{kge}) to the leading order in surface gravity.
\bea
V(r)&=&\frac13 l(l+1)\frac{{k_H}^2}{(\kappa R_H)^2\cosh^2(k_H r^*)}+B\frac{k_H^2 (k_Hr^*)^2}{\cosh^2(k_Hr^*)} \nonumber\\
&&+ O({k_H}^3).
\eea

As $B$ gets larger while keeping $l$ fixed, the second term makes this potential barrier effectively wider, more similar in shape to the rectangular potential barrier or the double delta-function barrier(see Fig. (\ref{V_extremal})),
where the resonances in transmission rate are well known \cite{Belchev:2011}. A resonance in transmission rate increases the transmission, so more
particles will penetrate the barrier, which in turn produces a spike in the greybody factor. This  qualitatively explains the resonance in the present scattering problem, as well as the tendency that the value of $B$
needed for resonance grows with $l$. In general, what kind of analytical potentials do or do not display the resonance in scattering is still an unsolved problem.

Although the analytical form of the effective potential $V(r)$ is not available for the case of small black holes, peaks are also expected when $B$ is large enough, from the above discussion. This is verified by Fig. (\ref{small}).

\begin{figure}[h]
  \centering
\includegraphics[height=0.27\textwidth,angle=0]{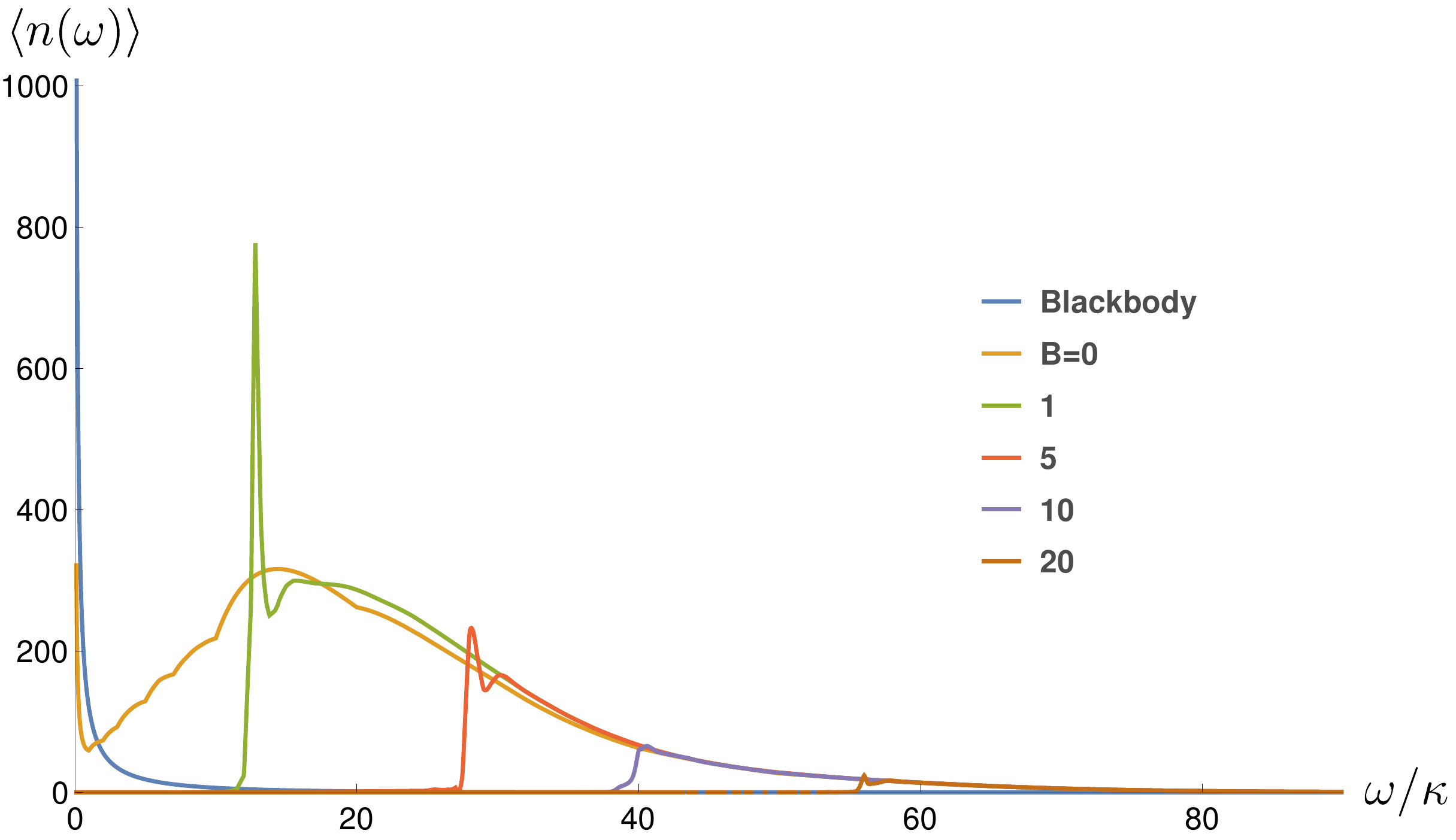}
\caption{The particle emission rate of small Schwarzschild-dS black holes ($\kappa R_H=0.01$) with different coupling with the background vector fields. The amplitudes of the black hole emission have all been amplified by $10^4$ times to be more visible.
 Different couplings are differentiated by different colors. Note that the greybody spectrum is always below the blackbody one before the artificial amplification.}
\label{n_w_small}
\end{figure}

\begin{figure}[h]
  \centering
\includegraphics[height=0.27\textwidth,angle=0]{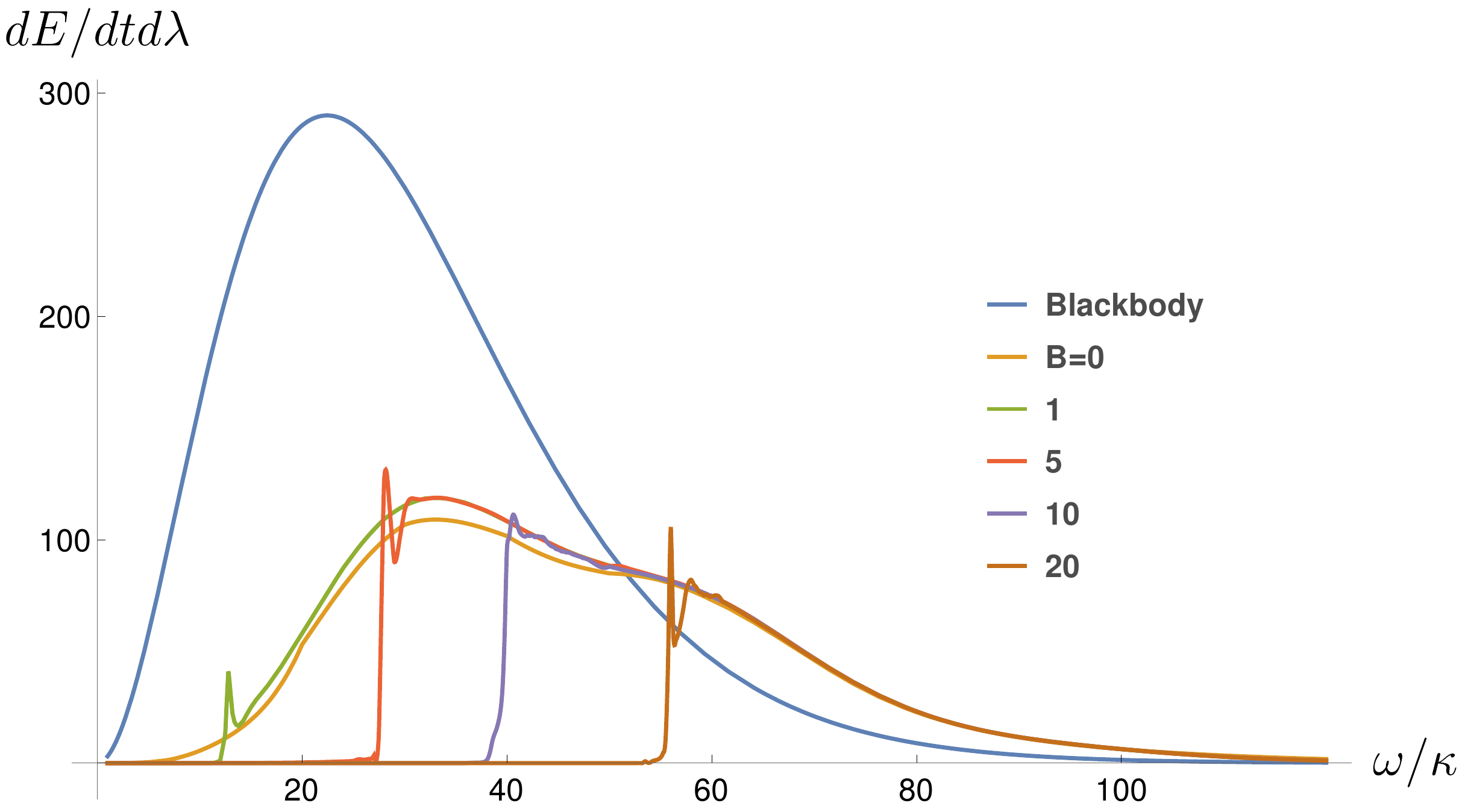}
\caption{The emission power spectrum of small Schwarzschild-dS black holes ($\kappa R_H=0.01$) with different coupling with the background vector fields. The amplitudes of the black hole emission have all been amplified by $10$ times to be more visible.
 Different couplings are differentiated by different colors. Note that the greybody spectrum is always below the blackbody one before the artificial amplification.}
\label{power_small}
\end{figure}

\begin{figure}[h]
  \centering
\includegraphics[height=0.27\textwidth,angle=0]{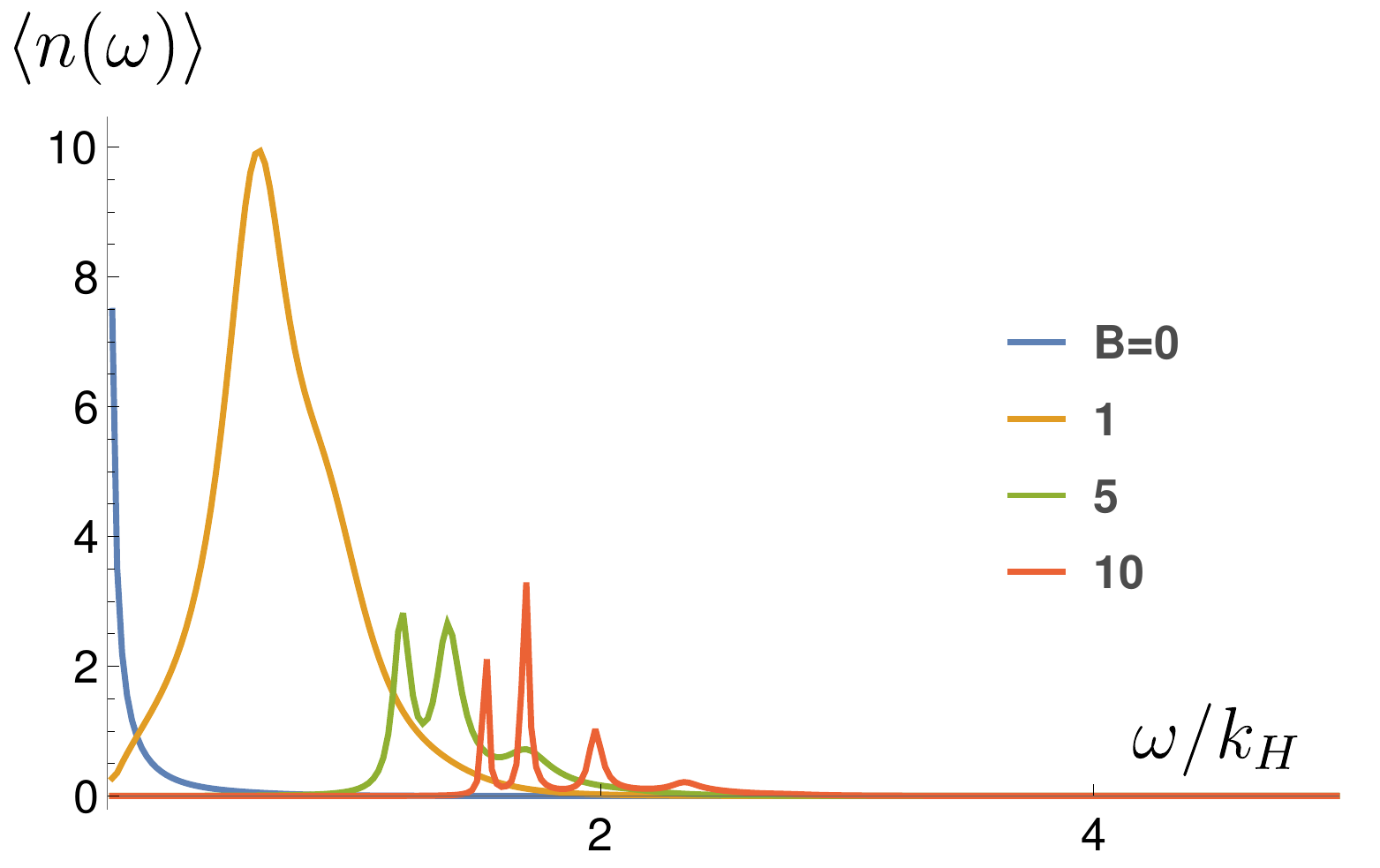}
\caption{The particle emission rate of near-extremal Schwarzschild-dS black holes with different coupling with the background vector fields. Different couplings are differentiated by different colors. The emission rates for $B=1, 5, 10$ are amplified by 500, 5000, 50000 times respectively, to be
more visible. Note that the $\langle n(\omega) \rangle$ decreases with increasing $B$ before this artificial amplification.}
\label{n_w_extremal}
\end{figure}

\begin{figure}[h]
  \centering
\includegraphics[height=0.27\textwidth,angle=0]{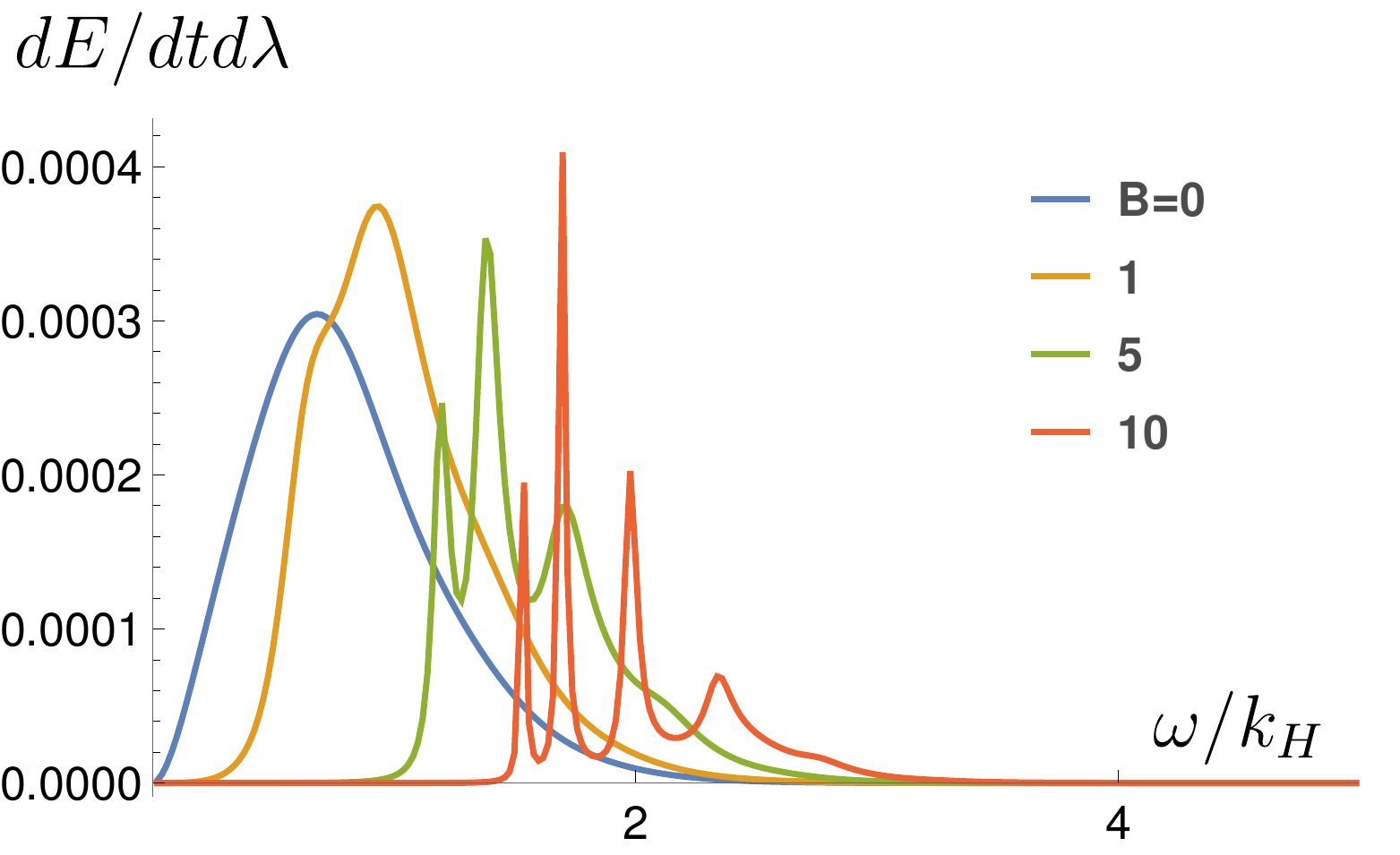}
\caption{The emission power spectrum of near-extremal Schwarzschild-dS black holes with different coupling with the background vector fields. The unit of $dE/dt d\lambda$ is $k_H^3$. Different couplings are differentiated by different colors. The power for $B=1, 5, 10$ are amplified by 2, 10, 50
times respectively, to be more visible. Note that the power gets more suppressed for larger $B$ before the artificial amplification.}
\label{power_extremal}
\end{figure}

For better comparison with the blackbody radiation and the case without coupling, we plot the particle emission rate $\langle n(\omega) \rangle$ and power spectrum $dE/dt d\lambda$, as shown in figures (\ref{n_w_small})(\ref{n_w_extremal})(\ref{power_small})(\ref{power_extremal}).
Although there are additional peaks in the greybody factors, which can raise the number and power emission, these peaks moves to higher frequencies as the coupling becomes stronger and so become less important. For example, for small black holes with $T_H\approx 8\kappa$ in Fig. (\ref{small})
for the $B=1$ case, only the peak in $\gamma_0(\omega)$ show up in figures (\ref{n_w_small})(\ref{power_small}), but the peaks in $\gamma_1(\omega), \gamma_2(\omega), \gamma_3(\omega)$ are all suppressed by the Boltzmann factor.

Similar things happen in the near-extremal case, where $T_H=\frac1{2\pi}k_H$. Spikes move to higher frequencies as $B$ gets larger, reducing their effects on the emission rate and spectrum. However, as seen from Fig. (\ref{extremal}), there are also "incidental" spikes. For example, $l$ from 0 to 8
all contribute spikes near $\omega/k_H=4$. So, conservatively, we only calculate the emission rate and spectrum for $B=0, 1, 5, 10$, using the greybody factors from $l$ up to $8$.

\section{Conclusions}

In this paper, we studied the greybody factor corrections to the Hawking radiation of black holes in massive gravity theory. We did the calculations numerically for small black holes, and analytically for the near-extremal ones (where the event and cosmological horizons are of the same order). Similarly, features of $\gamma_l(\omega)$ were found, with very small values at low frequencies while approaching unity at higher frequencies, and the main contribution to the radiated power comes from low-$l$ modes.

To the best of our knowledge, the first results on greybody factors of Schwarzschild-dS black holes (though in a different underlying context) were obtained by Kanti et al. \cite{Kanti:2005ja}. Also, in recent papers \cite{Kanti:2014dxa,Crispino:2013pya}, analytical form of the greybody factors for Schwarzschild-dS black holes were obtained, consistent with our results, by matching the solution forms (usually hypergeometric functions) in different regions of $r$. However, these approximations only hold either for small black holes, or for low energy modes. In contrast, we worked out the calculations for small black holes numerically, while for the near-extremal black holes analytically (exactly). This special case for the near-extremal black holes was found more than $30$ years ago for normal mode
calculations \cite{Ferrari:1984zz}, but remained unnoticed for computing the transmission coefficients.

Once we couple the test field to the the background {St\"uckelberg} fields, the situation becomes more complicated.
The background fields can be decomposed into two modes, of which the vector mode has $r$-dependence only, while the scalar mode has $t$-dependence. We consistently couple the test field with the vector mode only, which in turn drives the greybody factors to higher frequencies and introduces some non-trivial features (resonances) as explained in Section \ref{cbg}.

Recently, a solution for charged black holes in nonlinear massive gravity was found in \cite{Cai:2012db}. It would be interesting to find the greybody factors for such black holes and compare them with those found here.

\begin{acknowledgments}
This work was partially supported by the US National Science Foundation, under Grant No. PHY-1066278 and PHY-1417317.
\end{acknowledgments}

\end{document}